\documentclass[iop, revtex4-2]{emulateapj}

\usepackage{color}
\usepackage{courier}
\usepackage[caption=false]{subfig}

\def\simlt{\lower.5ex\hbox{$\; \buildrel < \over \sim \;$}}
\def\simgt{\lower.5ex\hbox{$\; \buildrel > \over \sim \;$}}

\def\gsim{\lower 2pt \hbox{$\, \buildrel {\scriptstyle >}\over
{\scriptstyle \sim}\,$}}
\def\lsim{\lower 2pt \hbox{$\, \buildrel {\scriptstyle <}\over
{\scriptstyle \sim}\,$}}

\def\deg{\ifmmode ^{\circ}
         \else $^{\circ}$\fi}
\def\pdeg{\ifmmode
           $\setbox0=\hbox{$^{\circ}$}\rlap{\hskip.11\wd0 .}$^{\circ}
     \else \setbox0=\hbox{$^{\circ}$}\rlap{\hskip.11\wd0 .}$^{\circ}$\fi}
     
\def\pc{\ifmmode \mathrm{pc} \else $\mathrm{pc}$ \fi}
\def\mpc{\ifmmode \mathrm{Mpc} \else $\mathrm{Mpc}$\fi}
\def\mpcthree{\ifmmode \mathrm{Mpc}^{-3} \else $\mathrm{Mpc}^{-3}$\fi}
\def\gpcthree{\ifmmode \mathrm{Gpc}^{-3} \else $\mathrm{Gpc}^{-3}$\fi}

\def\kelvin{\ifmmode \mathrm{K} \else {$\mathrm{K}$}\fi}
\def\kev{\ifmmode \mathrm{keV} \else $\mathrm{keV}$ \fi}

\def\lsun{\ifmmode {L_\odot} \else $L_\odot$\fi}
\def\msun{\ifmmode M_\odot \else $M_\odot$\fi}
\def\msunyr{\ifmmode M_\odot~\mathrm{yr}^{-1} \else $M_\odot~\mathrm{yr}^{-1}$\fi}

\def\cosi{\ifmmode {\cos\,i} \else $\cos\,i$\fi}

\def\heii{\ifmmode {\rm He{\sc ii}} \else He~{\sc ii}\fi}
\def\mgii{\ifmmode {\rm Mg{\sc ii}} \else Mg~{\sc ii}\fi}
\def\caii{\ifmmode {\rm Ca{\sc ii}} \else Ca~{\sc ii}\fi}
\def\ciii{\ifmmode {\rm C{\sc iii}]} \else C~{\sc iii}]\fi}
\def\civ{\ifmmode {\rm C{\sc iv}} \else C~{\sc iv}\fi}
\def\mgii{\ifmmode {\rm Mg{\sc ii}} \else Mg~{\sc ii}\fi}

\newcommand{\oiii}{{\sc [O~iii]}}

\def\teff{\ifmmode {T_{\rm eff}} \else $T_{\rm eff}$\fi}
\def\tmax{\ifmmode {T_{\rm max}} \else $T_{\rm max}$\fi}

\def\mbh{\ifmmode {M_{\rm BH}} \else $M_{\rm BH}$\fi}
\def\led{\ifmmode L_{\mathrm{Ed}} \else $L_{\mathrm{Ed}}$\fi}
\def\lbolflare{\ifmmode L_{\mathrm{bol,flare}} \else $L_{\mathrm{bol,flare}}$\fi}
\def\lagn{\ifmmode L_{\mathrm{agn}} \else $L_{\mathrm{agn}}$\fi}
\def\lbolagn{\ifmmode L_{\mathrm{bol,agn}} \else $L_{\mathrm{bol,agn}}$\fi}
\def\lbol{\ifmmode L_{\mathrm{bol}} \else $L_{\mathrm{bol}}$\fi}
\def\mdot{\ifmmode {\dot M} \else $\dot M$\fi}
\def\mdoto{\ifmmode {\dot{M}_0} \else  $\dot{M}_0$\fi}
\def\mdotf{\ifmmode {\dot{M}_\mathrm{flare}} \else  $\dot{M}_\mathrm{flare}$\fi}

\def\hnot{\ifmmode H_0 \else H$_0$ \fi}

\def\vkep{\ifmmode v_\mathrm{Kep} \else $v_\mathrm{Kep}$ \fi}
\def\vc{\ifmmode v_\mathrm{c} \else $v_\mathrm{c}$ \fi}

\def\vthree{\ifmmode v_{1000} \else $v_{1000}$ \fi}
\def\vrel{\ifmmode v_\mathrm{rel} \else $v_\mathrm{rel}$ \fi}
\def\vkick{\ifmmode v_\mathrm{kick} \else $v_\mathrm{kick}$ \fi}
\def\vkickz{\ifmmode v_{\mathrm{kick},z} \else $v_{\mathrm{kick},z} $ \fi}
\def\vkicky{\ifmmode v_{\mathrm{kick},y} \else $v_{\mathrm{kick},y} $ \fi}
\def\vchar{\ifmmode v_\mathrm{char} \else $v_\mathrm{char}$ \fi}
\def\eflare{\ifmmode E_\mathrm{flare} \else $E_\mathrm{flare}$ \fi}
\def\ekick{\ifmmode E_\mathrm{kick} \else $E_\mathrm{kick}$ \fi}
\def\ecoll{\ifmmode E_\mathrm{coll} \else $E_\mathrm{coll}$ \fi}
\def\ezero{\ifmmode E_\mathrm{0} \else $E_\mathrm{0}$ \fi}
\def\efac{\ifmmode \xi_\mathrm{E} \else $\xi_\mathrm{E}$ \fi}
\def\tqso{\ifmmode t_\mathrm{QSO} \else $t_\mathrm{QSO}$ \fi}
\def\tflare{\ifmmode t_\mathrm{flare} \else $t_\mathrm{flare}$ \fi}
\def\tzero{\ifmmode t_\mathrm{0} \else $t_\mathrm{0}$ \fi}
\def\tfac{\ifmmode \xi_\mathrm{t} \else $\xi_\mathrm{t}$ \fi}
\def\gfac{\ifmmode f_\mathrm{g} \else $f_\mathrm{g}$ \fi}
\def\lflare{\ifmmode L_\mathrm{flare} \else $L_\mathrm{flare}$ \fi}
\def\fflare{\ifmmode F_\mathrm{flare} \else $F_\mathrm{flare}$ \fi}
\def\nflare{\ifmmode N_\mathrm{flare} \else $N_\mathrm{flare}$ \fi}
\def\tshock{\ifmmode T_\mathrm{shock} \else $T_\mathrm{shock}$ \fi}
\def\rmin{\ifmmode R_\mathrm{1} \else $R_\mathrm{1}$ \fi}
\def\rmax{\ifmmode R_\mathrm{2} \else $R_\mathrm{2}$ \fi}
\def\rbound{\ifmmode R_\mathrm{b} \else $R_\mathrm{b}$ \fi}
\def\pbound{\ifmmode P_\mathrm{b} \else $P_\mathrm{b}$ \fi}
\def\mbound{\ifmmode M_\mathrm{b} \else $M_\mathrm{b}$ \fi}
\def\mbo{\ifmmode M_{\mathrm{b}0} \else $M_{\mathrm{b}0} $ \fi}
\def\ebo{\ifmmode E_{\mathrm{b}0} \else $E_{\mathrm{b}0} $ \fi}
\def\efinal{\ifmmode E_\mathrm{final} \else $E_\mathrm{final} $ \fi}
\def\tbound{\ifmmode t_\mathrm{b} \else $t_\mathrm{b}$ \fi}
\def\tagn{\ifmmode t_\mathrm{AGN} \else $t_\mathrm{AGN}$ \fi}
\def\torb{\ifmmode t_\mathrm{orb} \else $t_\mathrm{orb}$ \fi}
\def\tdf{\ifmmode t_\mathrm{df} \else $t_\mathrm{df}$ \fi}
\def\rlim{\ifmmode R_\mathrm{lim} \else $R_\mathrm{lim}$ \fi}
\def\vlim{\ifmmode v_\mathrm{lim} \else $v_\mathrm{lim}$ \fi}
\def\vphi{\ifmmode v_\phi \else $v_\phi$ \fi}
\def\mlim{\ifmmode M_\mathrm{lim} \else $M_\mathrm{lim}$ \fi}
\def\tlim{\ifmmode t_\mathrm{lim} \else $t_\mathrm{lim}$ \fi}
\def\llim{\ifmmode L_\mathrm{lim} \else $L_\mathrm{lim}$ \fi}
\def\fqso{\ifmmode f_\mathrm{QSO} \else $f_\mathrm{QSO}$ \fi}

\def\hbeta{\ifmmode \rm{H}\beta \else H$\beta$\fi}
\def\hbetan{\ifmmode \rm{H}\beta_{\rm n} \else H$\beta_{\rm n}$\fi}
\def\hgamma{\ifmmode \rm{H}\gamma \else H$\gamma$\fi}
\def\hdelta{\ifmmode \rm{H}\delta \else H$\delta$\fi}
\def\hepsilon{\ifmmode \rm{H}\epsilon \else H$\epsilon$\fi}
\def\hzeta{\ifmmode \rm{H}\zeta \else H$\zeta$\fi}
\def\halpha{\ifmmode \rm{H}\alpha \else H$\alpha$\fi}
\def\lalpha{\ifmmode \rm{Ly}\alpha \else Ly$\alpha$}

\def\dvhb{\ifmmode \Delta v_{\hbeta} \else $\Delta v_{\hbeta}$\fi}
\def\dvmg{\ifmmode \Delta v_{\rm{Mg}} \else $\Delta v_{\rm{Mg}}$\fi}

\def\muobs{\ifmmode {\mu_{o}} \else  $\mu_{o}$ \fi}
\def\cosi{\ifmmode {\mathrm{cos}\,i} \else $\mathrm{cos}\,i$\fi}

\def\teff{\ifmmode {T_{eff}} \else $T_{eff}$ \fi}
\def\tmax{\ifmmode {T_{max}} \else $T_{max}$ \fi}

\def\tauh{\ifmmode {\tau_{\rm H}} \else $\tau_{\rm H}$ \fi}

\def\yr{\ifmmode {\rm yr} \else  yr \fi}
\def\kms{\ifmmode \rm km~s^{-1}\else $\rm km~s^{-1}$\fi}
\def\cm{\ifmmode {\rm cm} \else  cm \fi}
\def\cmmitwo{\ifmmode \rm cm^{-2} \else $\rm cm^{-2}$\fi}
\def\cmmithree{\ifmmode \rm cm^{-3} \else $\rm cm^{-3}$\fi}
\def\cmps{\ifmmode \rm cm~s^{-1}\else $\rm cm~s^{-1}$\fi}
\def\cmpsps{\ifmmode \rm cm~s^{-2}\else $\rm cm~s^{-2}$\fi}
\def\kmps{\ifmmode \rm km~s^{-1}\else $\rm km~s^{-1}$\fi}
\def\kmpspmpc{\ifmmode \rm km~s^{-1}~Mpc^{-1} \else
    $\rm km~s^{-1}~Mpc^{-1}$\fi}
  
\def\gcmthree{\ifmmode \rm g~cm^{-3} \else $\rm g~cm^{-3}$\fi}
\def\gcmtwo{\ifmmode \rm g~cm^{-2} \else $\rm g~cm^{-2}$\fi}
   
\def\erg{\ifmmode {\rm erg} \else $\rm erg$ \fi}
\def\ergps{\ifmmode {\rm erg~s^{-1}} \else $\rm erg~s^{-1}$ \fi}
\def\ergcms{\ifmmode \rm erg~cm^{-2}~s^{-1} \else $\rm erg~cm^{-2}~s^{-1}$ \fi}
\def\ergcmshz{\ifmmode \rm erg~s^{-1}~cm^{-2}~Hz^{-1} \else $\rm
erg~cm^{-2}~s^{-1}~Hz^{-1}$ \fi}
\def\ergcmsa{\ifmmode \rm erg~cm^{-2}~s^{-1}~\AA^{-1} \else $\rm
erg~cm^{-2}~s^{-1}~\AA^{-1}$ \fi}
\def\ergshz{\ifmmode \rm erg s^{-1} Hz^{-1} \else
   $\rm erg s^{-1} Hz^{-1}$ \fi}

\def\lam{\ifmmode {\lambda} \else {$\lambda$} \fi}
\def\llam{\ifmmode {L_\lambda} \else  $L_\lambda$ \fi}
\def\lamLlam{\ifmmode \lambda L_{\lambda}(5100) \else {$\lambda L_{\lambda}(5100)$} \fi}
\def\nuLnu{\ifmmode \nu L_{\nu}(5100) \else {$\nu L_{\nu}(5100)$} \fi}
\def\ilam{\ifmmode {I_\lambda} \else  $I_\lambda$ \fi}
\def\flam{\ifmmode {F_\lambda} \else  $F_\lambda$ \fi}
\def\inu{\ifmmode {I_\nu} \else  $I_\nu$ \fi}
\def\fnu{\ifmmode {F_\nu} \else  $F_\nu$ \fi}
\def\bnu{\ifmmode {B_\nu} \else  $B_\nu$ \fi}

\def\msigma{\ifmmode M_{\sigma} \else $M_{\sigma}$\fi}
\def\mbulge{\ifmmode M_{\mathrm{bulge}} \else $M_{\mathrm{bulge}}$\fi}
\def\mgal{\ifmmode M_{\mathrm{gal}} \else $M_{\mathrm{gal}}$\fi}
\def\lgal{\ifmmode L_{\mathrm{gal}} \else $L_{\mathrm{gal}}$\fi}
\def\lbulge{\ifmmode L_{\mathrm{bulge}} \else $L_{\mathrm{bulge}}$\fi}
\def\mgalstar{\ifmmode M^*_{\mathrm{gal}} \else $M^*_{\mathrm{gal}}$\fi}

\def\mbhsigstar{\ifmmode M_{\mathrm{BH}} - \sigma_* \else $M_{\mathrm{BH}} - \sigma_*$ \fi}
\def\deltalogmbh{\ifmmode \Delta~{\mathrm{log}}~M_{\mathrm{BH}} \else $\Delta$~log~$M_{\mathrm{BH}}$\fi}

\def\sigstar{\ifmmode \sigma_* \else $\sigma_*$\fi}
\def\sigthree{\ifmmode \sigma_{\mathrm{[O~III]}} \else $\sigma_{\mathrm{[O~III]}}$\fi}
\def\sigtwo{\ifmmode \sigma_{\mathrm{[O~II]}} \else $\sigma_{\mathrm{[O~II]}}$\fi}
\def\signl{\ifmmode \sigma_{\mathrm{NL}} \else $\sigma_{\mathrm{NL}}$\fi}
\def\wthree{\ifmmode {\rm FWHM({[O~III]})} \else $FWHM({[O~III]})$ \fi}
\def\wtwo{\ifmmode {\rm FWHM({[O~II]})} \else $FWHM({[O~II]})$ \fi}
\def\mthree{\ifmmode M_{\mathrm [O~III]} \else $M_{\mathrm [O~III]}$ \fi}
\def\mtwo{\ifmmode M_{\mathrm [O II]} \else $M_{\mathrm [O II]}$ \fi}

\def\lbreak{\ifmmode L_{\mathrm{break}} \else $L_{\mathrm{break}}$\fi}
\def\lcut{\ifmmode L_{\mathrm{cut}} \else $L_{\mathrm{cut}}$\fi}

\shortauthors{Smith, Koss, Mushotzky, Wong, Shimizu, Ricci \& Ricci}
\shorttitle{Kiloparsec-scale radio structures and feedback}

\begin{document}

\title{Significant Suppression of Star Formation in Radio-Quiet AGN Host Galaxies with Kiloparsec-Scale Radio Structures}

\author{Krista Lynne Smith\altaffilmark{1,2}, Michael Koss\altaffilmark{3}, Richard Mushotzky\altaffilmark{4}, O. Ivy Wong\altaffilmark{5,6}, T. Taro Shimizu\altaffilmark{7}, Claudio Ricci\altaffilmark{8,9,10} \& Federica Ricci \altaffilmark{11}}

\altaffiltext{1}{Department of Physics, Southern Methodist University, Dallas, TX 75205, USA; kristas@smu.edu}
\altaffiltext{2}{KIPAC at SLAC, Stanford University, Menlo Park, CA 94025, USA}
\altaffiltext{3}{Eureka Scientific, 2452 Delmer Street Suite 100, Oakland, CA 94602-3017, USA}
\altaffiltext{4}{Department of Astronomy and Joint Space-Science Institute, University of Maryland, College Park, MD 20742, USA}
\altaffiltext{5}{CSIRO Astronomy \& Space Science, PO Box 1130, Bentley, WA 6102, Australia}
\altaffiltext{6}{ICRAR-M468, University of Western Australia, Crawley, WA 6009, Australia}
\altaffiltext{7}{Max-Planck-Institut f\"{u}r extraterrestrische Physik, Postfach 1312, 85741, Garching, Germany}
\altaffiltext{8}{N\'ucleo de Astronom\'ia de la Facultad de Ingenier\'ia, Universidad Diego Portales, Av. Ej\'ercito Libertador 441, Santiago, Chile}
\altaffiltext{9}{Kavli Institute for Astronomy and Astrophysics, Peking University, Beijing 100871, China}
\altaffiltext{10}{George Mason University, Department of Physics \& Astronomy, MS 3F3, 4400 University Drive, Fairfax, VA 22030, USA}
\altaffiltext{11}{Instituto de Astrof\'isica and Centro de Astroingenier\'ia, Facultad de F\'isica, Pontificia Universidad Católica de Chile, Casilla 306, Santiago 22, Chile}

\begin{abstract}

We conducted 22~GHz 1\arcsec~JVLA imaging of 100 radio-quiet X-ray selected AGN from the \emph{Swift}-BAT survey. We find AGN-driven kiloparsec-scale radio structures inconsistent with pure star formation in 11 AGN. The host galaxies of these AGN lie significantly below the star-forming main sequence, indicating suppressed star formation. While these radio structures tend to be physically small compared to the host galaxy, the global star formation rate of the host is affected. We evaluate the energetics of the radio structures interpreted first as immature radio jets, and then as consequences of an AGN-driven radiative outflow, and compare them to two criteria for successful feedback: the ability to remove the CO-derived molecular gas mass from the galaxy gravitational potential and the kinetic energy transfer to molecular clouds leading to $v_\mathrm{cloud} > \sigma_*$. In most cases, the jet interpretation is insufficient to provide the energy necessary to cause the star formation suppression. Conversely, the wind interpretation provides ample energy in all but one case. We conclude that it is more likely that the observed suppression of star formation in the global host galaxy is due to ISM interactions of a radiative outflow, rather than a small-scale radio jet.

\end{abstract}


\section{Introduction}
\label{sec:intro}

While powerful, megaparsec-scale relativistic jets have long been a classical signature of quasars and radio galaxies, and apparently have profound effects on their hosts and cluster environments \citep{Fabian2012}, they occur in only a relatively small fraction of AGN \citep{Kellermann1989}. In the more numerous, less luminous population of Seyfert galaxies, radio emission on kiloparsec scales is now recognized to be quite common: once the radio emission from star formation is accounted for, many radio-quiet Seyferts exhibit compact outflows, linear jet-like radio morphologies, S-shaped radio morphologies, and other nuclear structures \citep{Colbert1996, Gallimore2006, Singh2015, Jarvis2019}. 

Such kiloparsec-scale radio structures (KSRs) may be analogous to the more powerful Fanaroff-Riley Type I and II jets seen in radio-loud sources, but are considerably less powerful, slower, and span $1-10$~kpc instead of megaparsecs. It is possible that interaction with the host interstellar medium (ISM) frustrates the jet and prevents it from reaching the spatial extent of its more powerful cousins \citep{ODea1991, Whittle2004}; although this frustration may be temporary. If the jets are young, they may eventually break free of the nuclear ISM confinement \citep{ODea1998, Bicknell2018}. 

On the other hand, these radio structures may not be actual AGN jets at all. Instead, they  may be caused by radiatively-driven AGN outflows interacting with the host ISM, causing shocks that result in synchrotron emission. This idea was first put forward by \citet{Stocke1992}, and is supported by an observed correlation of \oiii~ velocity with radio power and total infrared luminosity \citep{Mullaney2013,Zakamska2014}. Such an interaction can result in a bimodal, pseudo-jet radio morphology: although the wind is generally assumed to initially propagate off the accretion disk in a spherically symmetric fashion \citep[e.g.,][]{King2010}, in a disk galaxy it will move more easily perpendicular to the gas disk along the path of least resistance, which results in cone- or bubble-like structures on either side of the nucleus. Radio emission fills these structures, a phenomenon observed in several nearby Seyferts \citep{Cecil2001, Hota2006, Croston2008,Congiu2017}. The effect that such AGN-driven outflows would have on star formation in the host is not clear; they could indeed conform to the usual negative feedback  models \citep[e.g., ][]{Hopkins2006}, or could involve positive feedback, especially along the edges of the outflows \citep[e.g., ][]{Silk2013, Zubovas2013}; see the review of many modes of feedback by \citet{Morganti2017}. 

It \emph{is} clear that there must be some regulatory communication between the central supermassive black hole and the host galaxy on larger scales, as evidenced by the apparent ubiquity of the correlation between the mass of a galaxy's central supermassive black hole and the stellar velocity dispersion of its bulge, known as M-$\sigma_*$ \citep{Gebhardt2000, Ferrarese2000, Gultekin2009}, and other empirical relations. The energy output provided by occasional active accretion by the central black hole is a primary candidate for providing the necessary regulation in a process commonly referred to as ``AGN feedback."

Host galaxies of AGN have been observed to lie in the otherwise sparsely-populated region between the blue, star forming sequence and the ``red and dead" ellipticals on color-mass plots \citep[e.g.,][]{Nandra2007, Schawinski2009}, interpreted as evidence that AGN are quenching the star formation. If this is true, then AGN host galaxies should have suppressed star formation rates compared to other galaxies of similar stellar mass. There is a well-studied linear relationship between the total stellar mass ($M_*$) of normal star-forming galaxies and their star formation rates (SFRs), known as the ``star forming main sequence" \citep{Brinchmann2004, Salim2007, Rodighiero2010}. If a galaxy has suppressed star formation for a given stellar mass, it will fall below this sequence. Many of the first main sequence (MS) studies to include AGN did not find that they lay systematically below this relation, despite expectations. Instead, AGN were found to lie mainly on the MS \citep{Mullaney2012, Rosario2013}, supporting a paradigm in which global star formation is uncoupled from nuclear activity. In this case, any global correlations of black hole mass and host properties must be the result of co-evolution or due to external stimuli that enhance both black hole growth and star formation. In a sample of 1000 X-ray selected AGN in the COSMOS field, \citet{Bongiorno2012} find evidence that AGN activity and star formation are likely triggered simultaneously, but find no evidence that AGN have a meaningful effect on the star formation rates of their hosts.  

On the other hand, many more recent studies have found that AGN hosts do indeed fall below the MS \citep{Salim2007,Ellison2016}, and that their loci on the SFR-$M_*$ diagram are nearly perpendicular to the main sequence \citep{Leslie2016}, concluding that quenching by AGN plays an important role in the evolutionary path of galaxies. \citet{McPartland2019} found that AGN do lie below the main sequence of star formation, but that this is a sensitive function of total galaxy mass. A recent study by \citet{Masoura2018} found a more complex situation, in which the effect of the AGN on the host (quenching vs. enhancing star formation) depended upon whether the AGN was above or below the main sequence of star formation, giving the AGN a powerful regulatory effect. Most recently, \citet{Stemo2020} found that AGN host galaxies reside below the main sequence, with X-ray selected AGN being further below than an IR-selected sample. 

The 70-month \emph{Swift}-BAT All-Sky Survey \citep{Baumgartner2013} was conducted in the ultra hard 14-195~keV X-ray band, and consists of over 1000 sources, approximately 700 of which are AGN. This AGN sample is highly unbiased compared to other surveys with respect to black hole mass, accretion rate, and obscuration \citep{Koss2017}. The host galaxies of the BAT AGN sample differ markedly from more biased samples selected at other wavelengths, with a large number of massive spirals ($>10^{10} M_\odot$) and mergers compared to inactive comparison samples \citep{Koss2010, Koss2011}.
\citet{Shimizu2015} calculated the SFRs and stellar masses of a subset of BAT AGN with \emph{Herschel} imaging to determine that nearby ($z<0.5$) ultra-hard X-ray selected AGN lay firmly between star-forming galaxies on the main sequence and quiescent galaxies far below, in the region sometimes referred to as the ``green valley." 

For the past four years we have been surveying the same subset of nearby, \emph{Herschel}-observed BAT AGN with the JVLA, resulting in our current atlas of 1\arcsec~ resolution 22~GHz radio imaging of 100 BAT AGN \citep{Smith2016, Smith2020a}. In this paper, we focus our attention on 11 AGN that exhibit jet-like kiloparsec-scale radio structures. Six of these were previously known to have such structures: 2MASX~J0423+0408 \citep{Beichman1985}; MCG+08-11-011 \citep{Wilson1980}; NGC~2110 \citep{Nagar1999}; NGC~3516 \citep{Miyaji1992}, although previous work saw this object as only one-sided where our observations seea two-sided structure); NGC~5548 \citep{Mathur1995}; and NGC~5728 \citep{Durre2018}. The other five targets are newly-discovered radio morphologies. Despite the typically small size of the radio structures compared to the host galaxies, and the fact that the sample is largely radio-quiet \citep{Smith2020a}, the global star formation of the hosts is significantly suppressed. For each object, we discuss whether the observed suppression is energetically likely to result from feedback analogous to classical relativistic jets or outflow interaction scenarios, and the implications for the nature of the KSRs.

In Section~\ref{sec:obs}, we discuss the properties of the sample and describe the radio observations, reduction techniques, and existing auxiliary data. Section~\ref{sec:mstarsfr} describes the calculation of the stellar mass and star formation rates and where our host galaxies lie with respect to the star-forming main sequence. In Section~\ref{sec:energetics} we perform estimations of the jet power and compare it to the energy required for star formation suppression. We then evaluate the energetics assuming the radio structure is evidence of a radiative outflow in Section~\ref{sec:wind}. Results are discussed in Section~\ref{sec:discussion}, and we summarize our conclusions in Section~\ref{sec:conclusions}.

When calculating luminosities from fluxes and spectroscopic redshifts, we assume a cosmology with the following parameters: $H_0 = 69.6$~km s$^{-1}$~Mpc$^{-1}$, $\Omega_M = 0.286$, and $\Omega_\Lambda = 0.714$ \citep{Bennett2014}.
\section{Sample and Data Reduction}
\label{sec:obs}

\subsection{Radio Survey Data}
\label{sec:radio_obs}
 
We obtained K-band (22~GHz) continuum observations of 100 nearby ($0.003< z< 0.049$) radio-quiet AGN from the \emph{Swift}-BAT survey with the JVLA in C-configuration, which has an angular resolution of $\sim1$\arcsec. The sample is the same \emph{Herschel}-observed subset of the BAT AGN used by \citet{Shimizu2017} to study star formation properties. The data were processed using the Common Astronomy Software Applications package  (CASA, \citealp{McMullin2007}). The detailed data reduction techniques used in the survey are described in \citet{Smith2020a}. Briefly, the survey was conducted in three campaigns between 2013 and 2017, consisting of observing blocks of 2-3 images each. Blocks began with X- and K-band attenuation scans and flux and bandpass calibrations with 3C~48, 3C~138, 3C~286, or 3C~147. Science integration times ranged between 3 and 10 minutes, with a typical 1$\sigma$~sensitivity of $\sim16~\mu$Jy per beam. After inspection and flagging to remove RFI banding, the final images underwent a 10,000 iteration CLEAN with a 0.03 mJy threshold. For objects with peak flux densities exceeding 1~mJy, we performed phase self-calibration on the visibility data.

This paper focuses on the 11 targets that have 22~GHz radio morphologies indicative of kiloparsec-scale AGN-driven jets or outflows. In the larger sample, 96/100 objects were robustly detected at 22~GHz. Of these, 55 were unresolved at the 1\arcsec~resolution of the survey, 30 exhibited extended star formation morphologies, and 11 exhibited jet-like morphology. Full classification information can be found in \citet{Smith2020a}. Two criteria were required for the present sample: 1) visual inspection indicating a linear or jet-like morphology, and 2) an excess of radio flux above that predicted from the SFR calculated from the infrared \emph{Herschel} data based on the far-IR / radio correlation for star forming regions \citep{Condon1992}. The two exceptions are NGC~3516 and NGC~5728: after subtracting the AGN core, NGC~5728 has slightly less radio emission than expected from global star formation, and  NGC~3516 almost exactly meets the expectation. We have included them because the observed radio structures have been established as AGN-related structures, and not star formation, in previous studies \citep{Wrobel988, Veilleux1993, Durre2018}.

These sources are listed in Table~\ref{t:tab1}. Figure~\ref{fig:hosts} shows the radio maps superimposed upon optical images of the host galaxies. The radio maps alone are given in either \citet{Smith2016} or \citet{Smith2020a}.


\begin{figure*}
\setlength{\tabcolsep}{1pt}
\begin{tabular}{ll}
  \includegraphics[width=80mm]{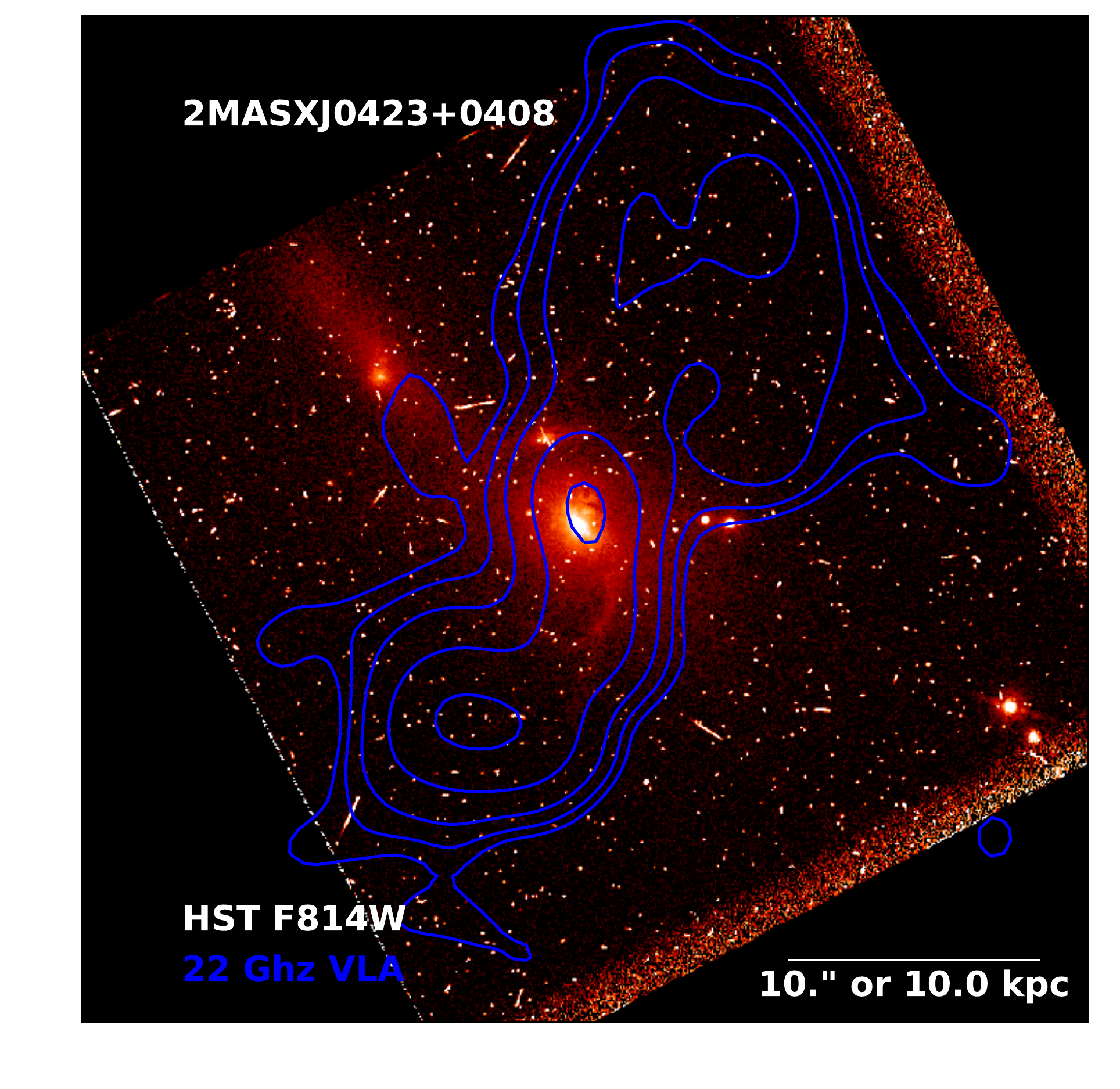} &   \includegraphics[width=80mm]{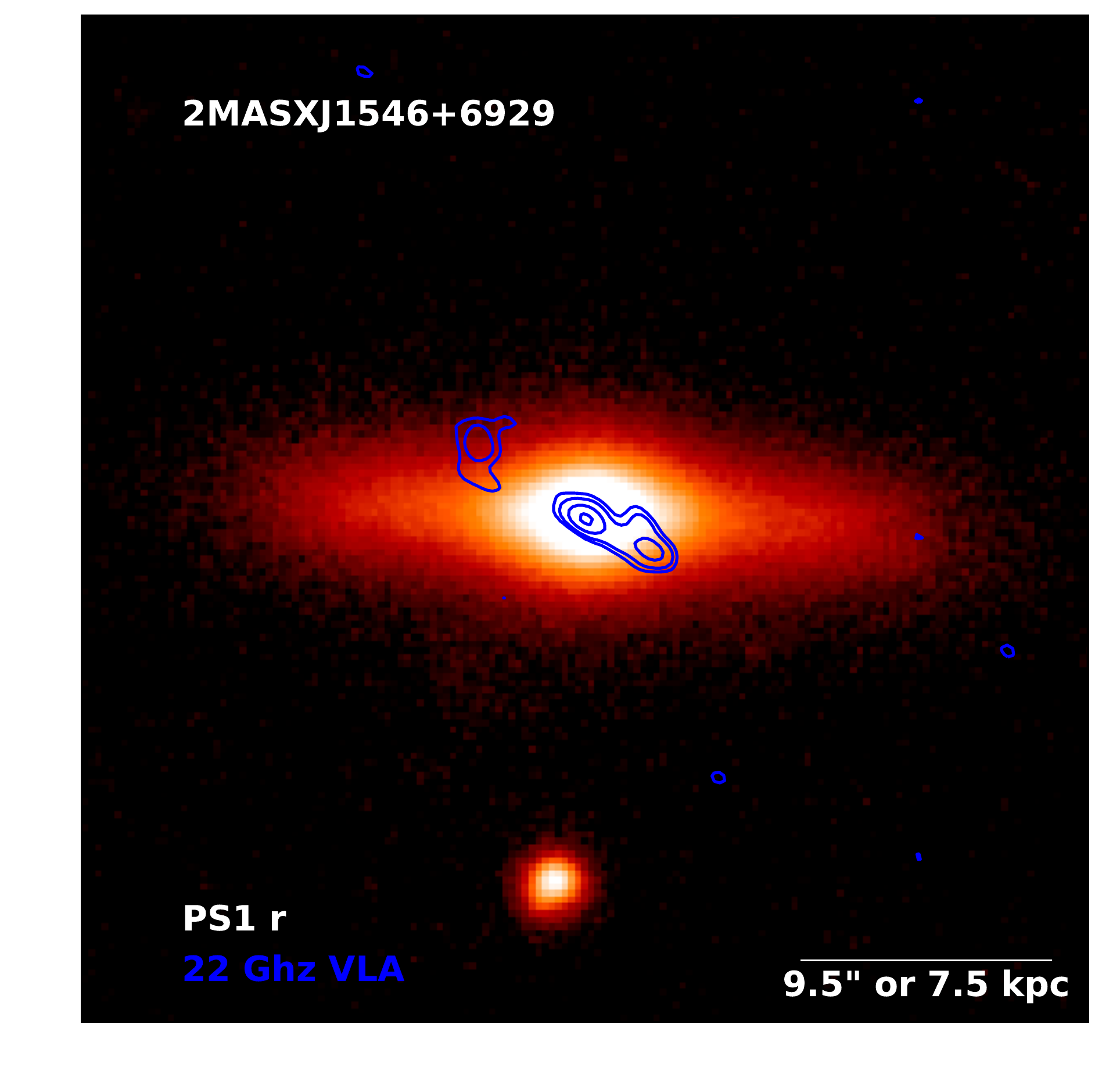} \\ \includegraphics[width=80mm]{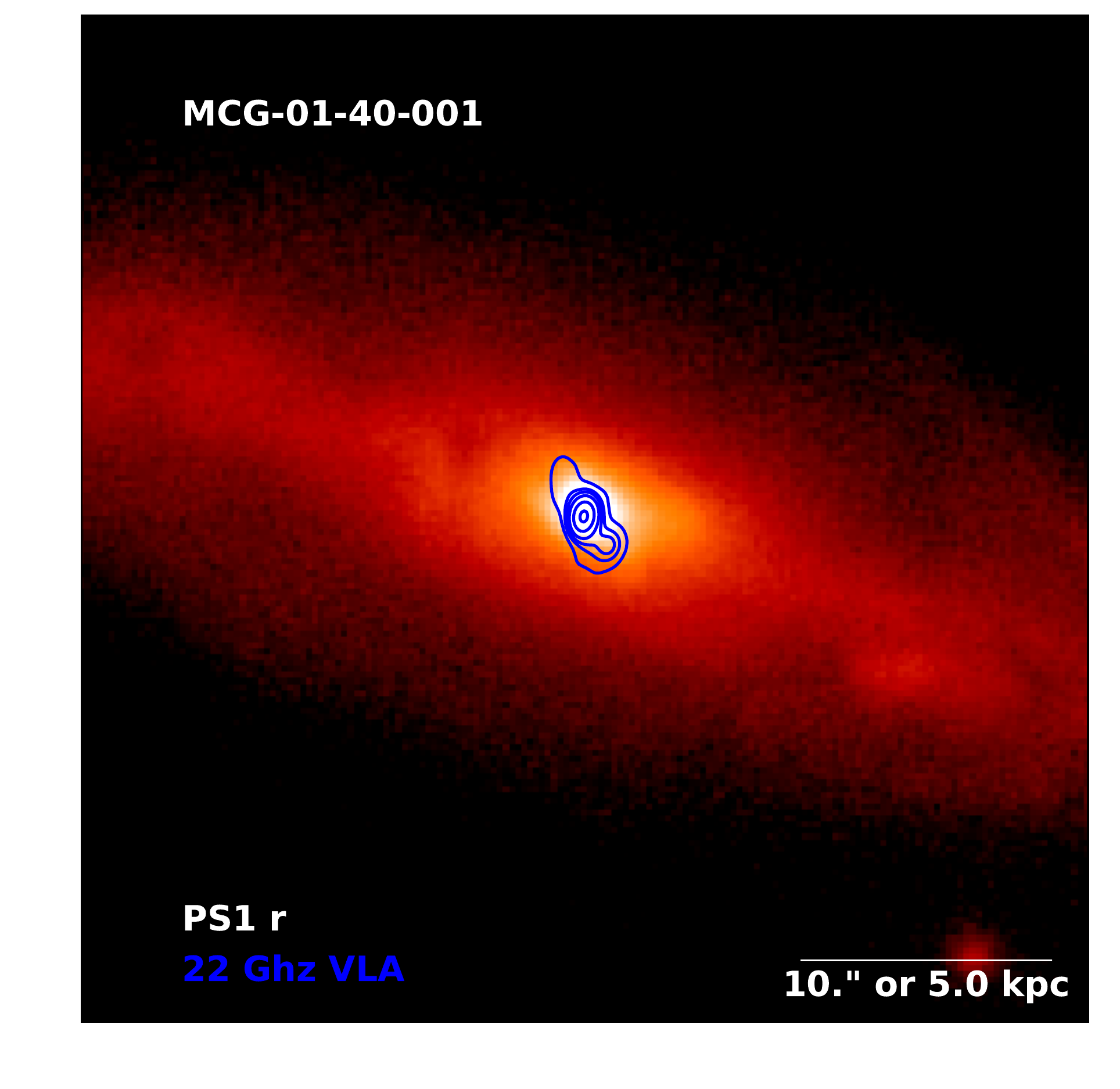} & \includegraphics[width=80mm]{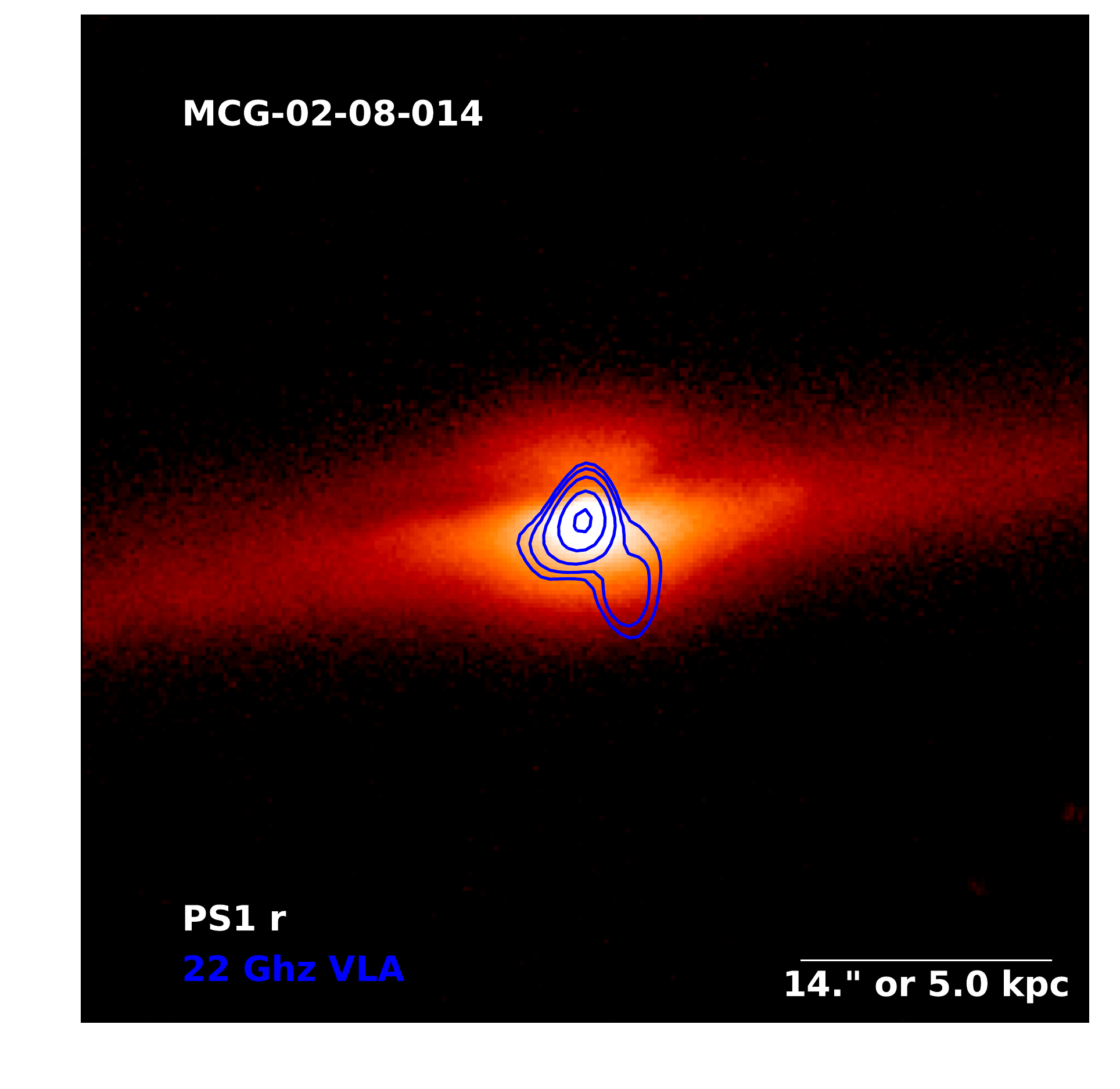} \\
 \includegraphics[width=80mm]{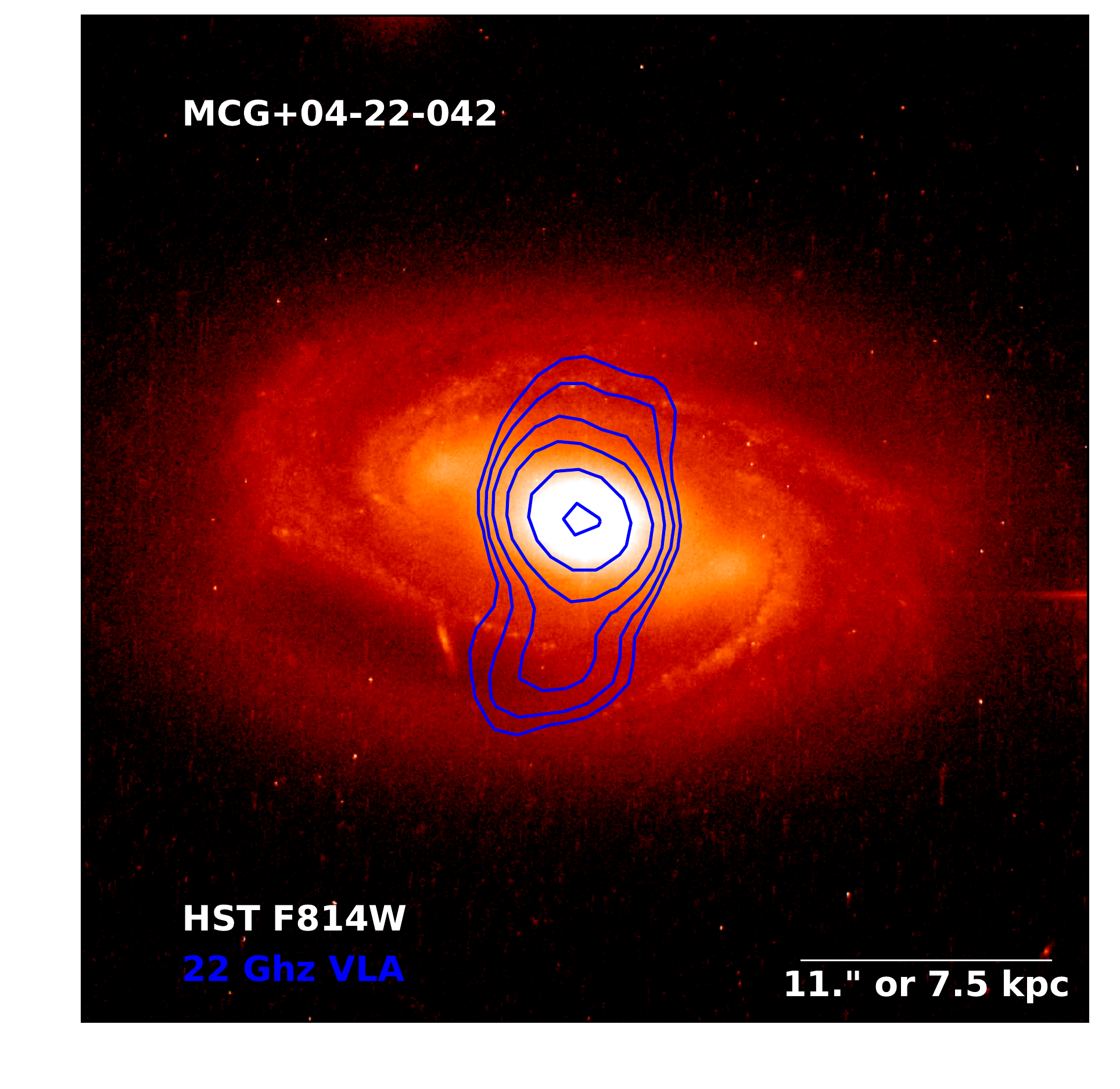} &   \includegraphics[width=80mm]{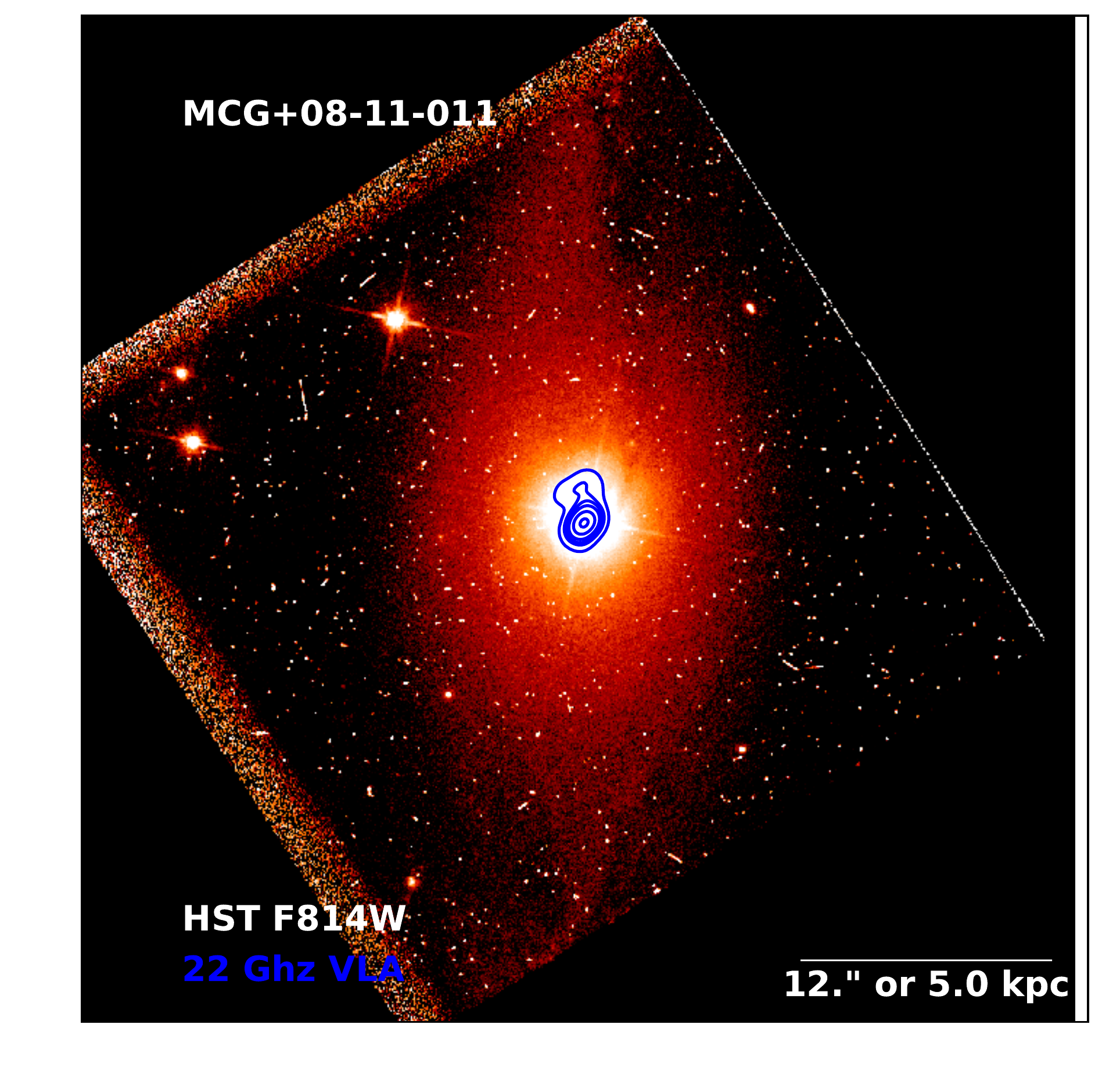} \\ 
\end{tabular}
\caption{Optical images of host galaxies with radio contours overlaid in blue. The contours are shown at 90, 50, 25, 15, and 10\% of the peak flux density, except in the case of NGC~3516 which is shown down to 3\%. The images are archival, from either the Hubble Space Telescope (HST) or the PanSTARRS survey \citep{Chambers2016}, as indicated in the bottom left of each image.
\label{fig:hosts}}
\end{figure*}

\begin{figure*}\ContinuedFloat
\begin{tabular}{ll}
\includegraphics[width=80mm]{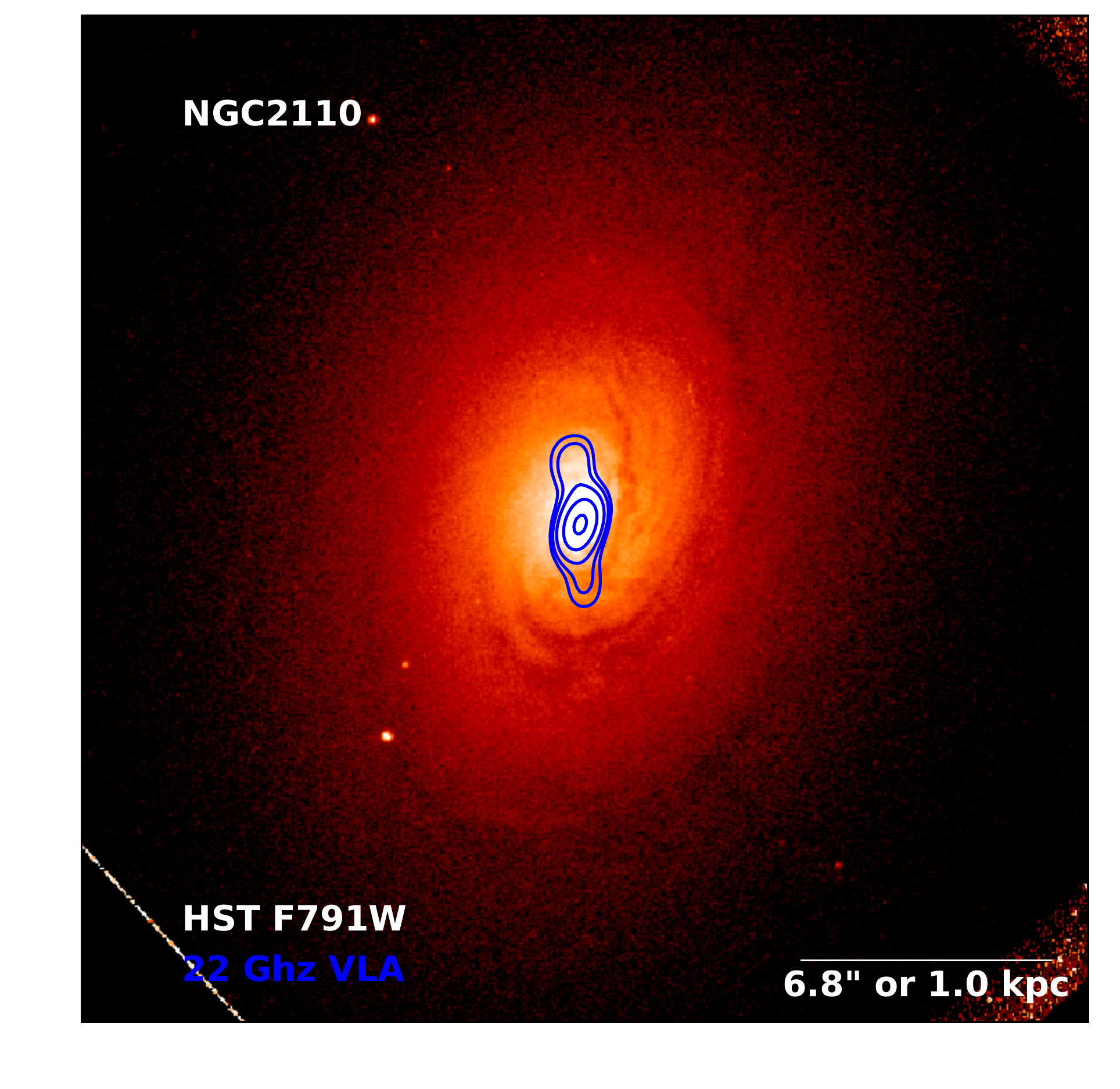} &
 \includegraphics[width=80mm]{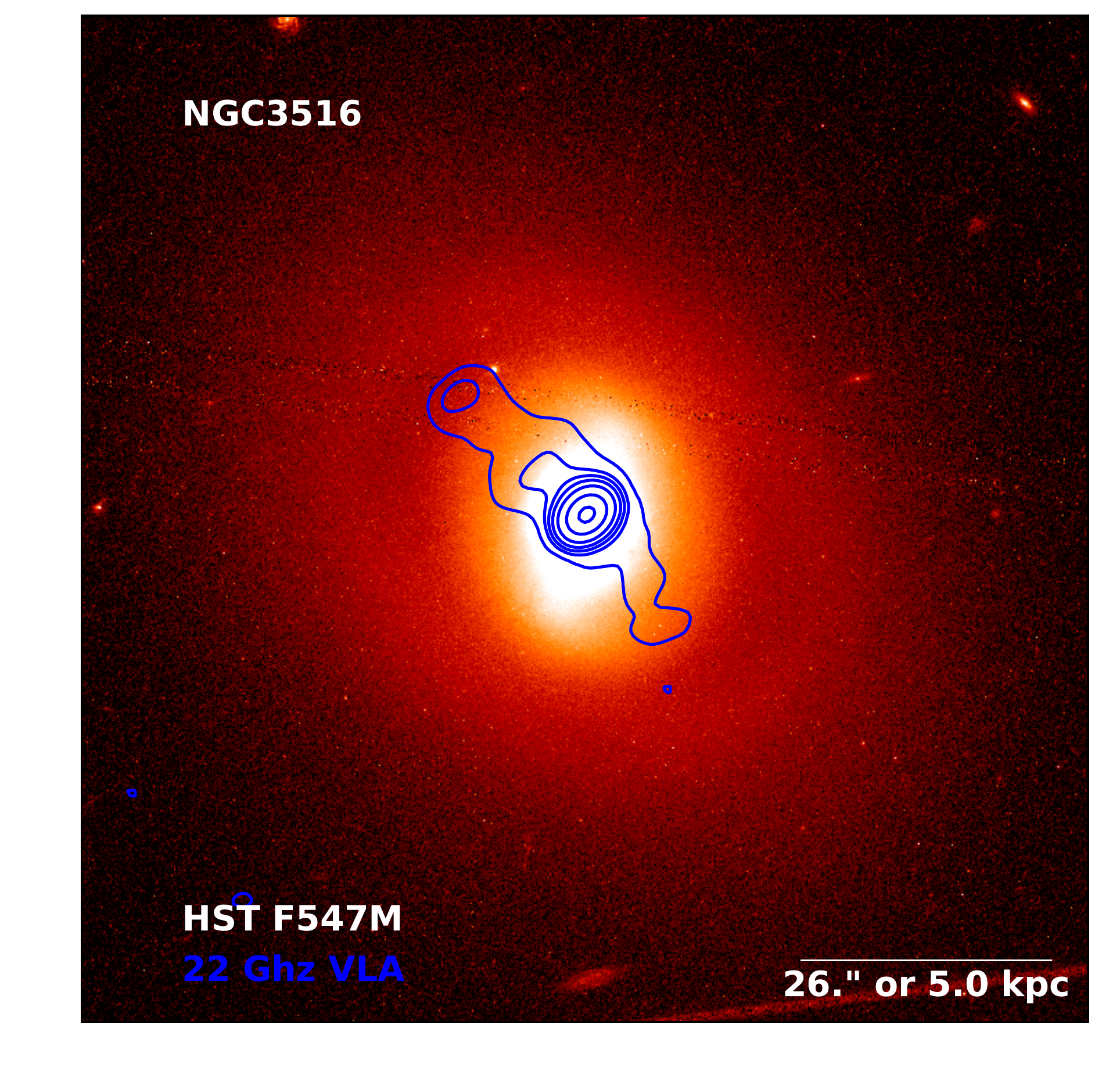} \\
 \includegraphics[width=80mm]{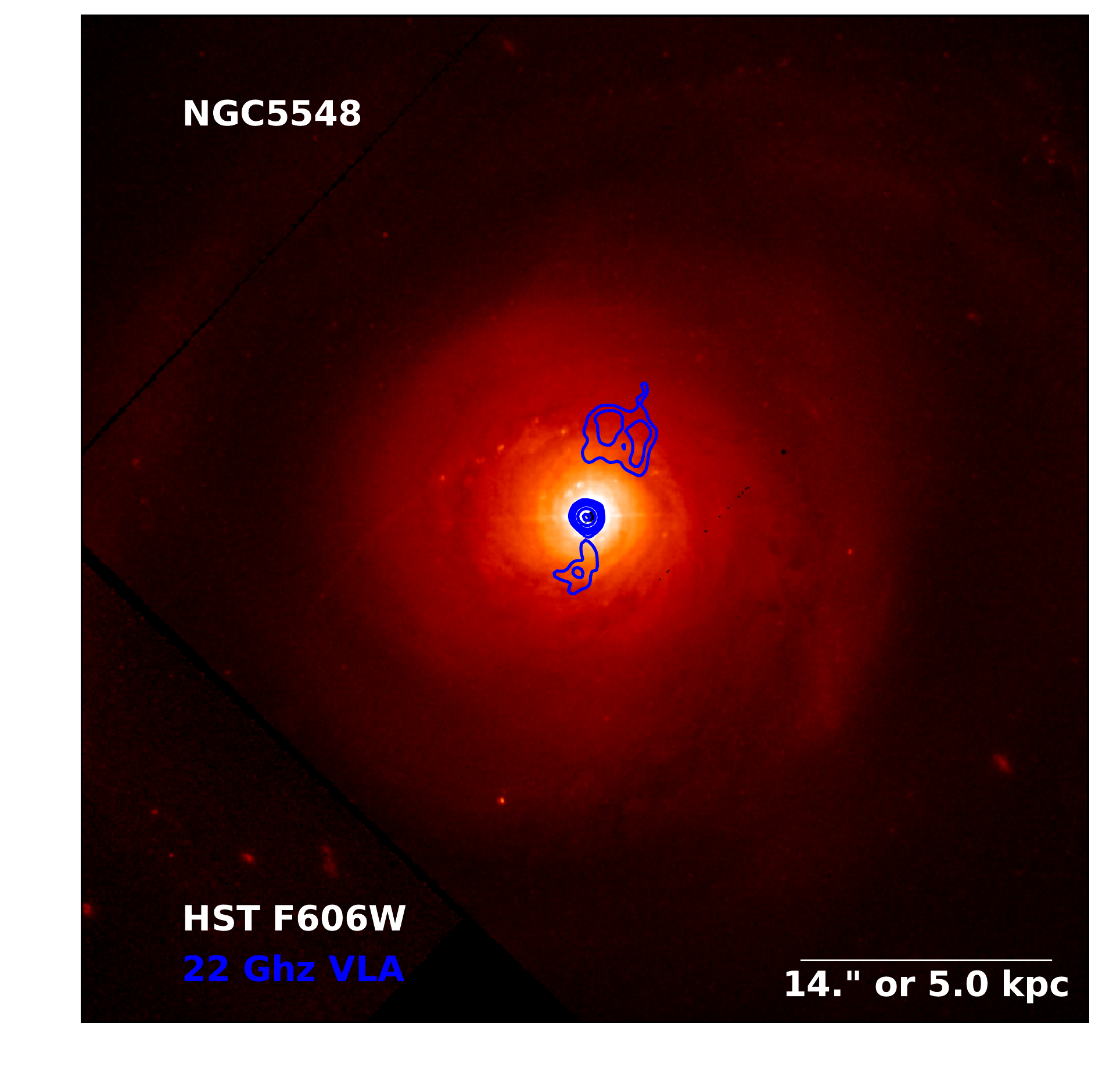} &   \includegraphics[width=80mm]{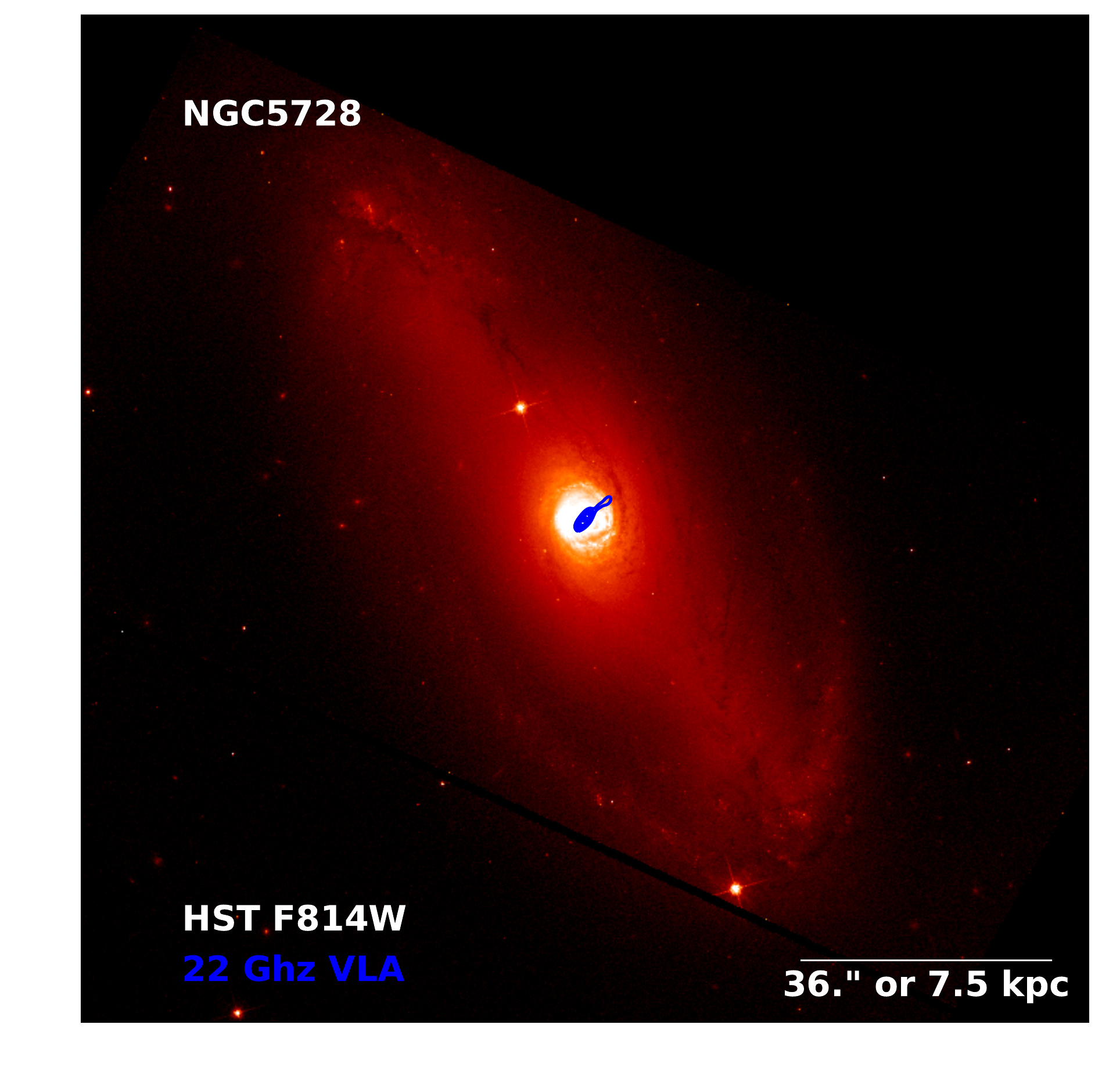} \\ \includegraphics[width=80mm]{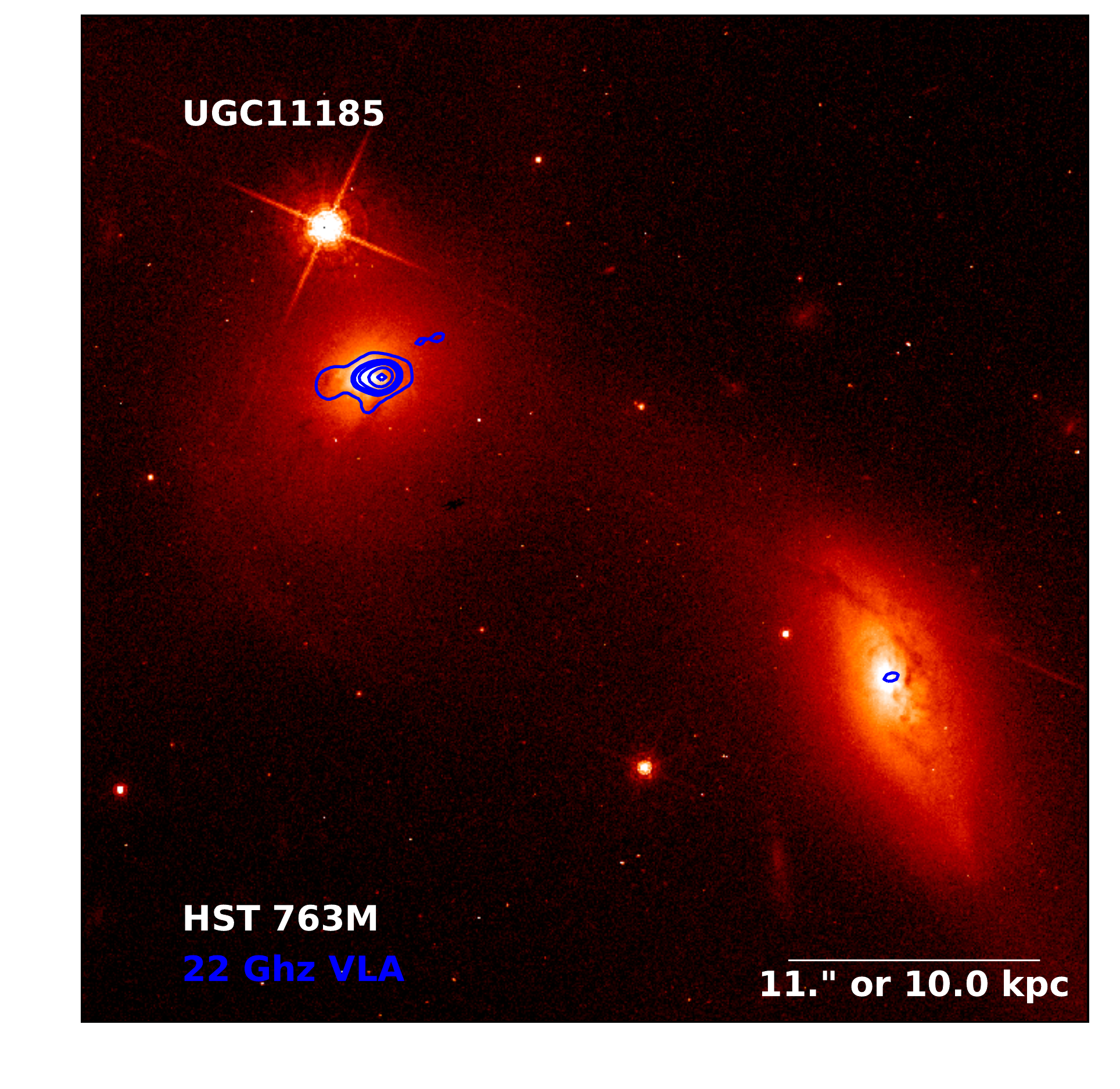} &
 \\
 \\
\end{tabular}
\caption{Continued from previous page.}
\end{figure*}

We note that we also recover several kiloparsec-scale radio structures from the literature \citep{Gallimore2006} that are likely to originate from star formation, including NGC~3227, NGC~3786, NGC~2992 and NGC~4388. They do not meet our jet criteria, as the observed 22~GHz emission does not far exceed that predicted from the infrared star formation component (see Section~\ref{sec:mstarsfr}). Throughout the manuscript, we use the term ``KSR" to refer to our targets with extended radio emission requiring a significant AGN component and with jet-like morphology, excluding these star-formation dominated objects. 

Finally, four objects from our larger sample have compact 22~GHz morphology, precluding our selecting them as KSR sources, but have known jets in lower frequency observations from the literature: Arp~102B, MCG-01-24-012, NGC~1052, and NGC~3718. \citet{Smith2020a} calculated the \citet{Kellermann1989} radio loudness criterion, $R$,  for these objects, where $R\sim10$ is traditionally considered the boundary between radio-loud and radio-quiet objects. Arp~102B ($R$ = 44) is radio-loud; NGC~1052 ($R=9$) and NGC~3718 ($R=2.4$) are marginally radio-loud, sometimes referred to as radio-intermediate, and MCG-01-24-012 ($R=0.12$) is radio-quiet. Arp~102B and NGC~1052 have elliptical host morphologies, distinct from all of the other targets in our sample. Because we perform all of our energetics calculations upon the observed extended 22~GHz emission, these objects are not included to retain consistency. We do include them graphically when considering the main sequence, however, for the curious reader.

\begin{deluxetable*}{lcccccccc}
 \tablewidth{0pt}
 \tablecaption{BAT AGN with Kiloparsec-Scale Radio Structures: Physical Parameters\label{t:tab1}}
 \tablehead{
 \colhead{Name} &
 \colhead{$z$} &
 \colhead{Sy} &
 \colhead{Host} &
 \colhead{log $M_*$} &
 \colhead{log SFR} &
  \colhead{$S_{\nu,\mathrm{22GHz}}$} &
  \colhead{$S_{\nu,\mathrm{5GHz}}$} &
  \colhead{$S_{\nu,\mathrm{1.4GHz}}$}\\
 \colhead{ } &
 \colhead{ } &
 \colhead{ Type } &
 \colhead{Morph. } &
 \colhead{ \footnotesize $M_{\odot}$ } &
 \colhead{ \footnotesize $M_{\odot}$ yr$^{-1}$} &
 \colhead{ \footnotesize mJy } &
 \colhead{ \footnotesize mJy } &
 \colhead{ \footnotesize mJy }\\
 \colhead{\tiny (1)} &
 \colhead{\tiny (2) } &
 \colhead{ \tiny (3) } &
 \colhead{ \tiny (4) } &
 \colhead{ \tiny (5) } &
 \colhead{ \tiny (6) } &
 \colhead{ \tiny (7)} &
 \colhead{\tiny (8) } &
  \colhead{\tiny (9) }\\
}

 \startdata 
2MASX J04234080+0408017      	&	0.045	&	2	&	Sb		&	10.22	&	0.64$^{+0.11}_{-0.19}$	&	5.90$\pm0.31$ & 16.64$\pm$2.07 & 40.57$\pm$2.07	\\
2MASX J15462424+6929102      	&	0.038	&	2	&	S0		&	10.78	&	-0.06$^{+0.16}_{-0.40}$	&	1.48$\pm0.07$ & 4.18$\pm$ 0.51 & 10.81$\pm$0.51	\\
MCG-01-40-001		&	0.023	&	2	&	Sab		&	10.88	&	0.54$^{+0.05}_{-0.06}$	&	23.00$\pm{0.15}$ & 64.89$\pm$7.34 & 158.17$\pm$ 7.34	\\
MCG-02-08-014		&	0.017	&	2	&	Sab		&	10.36	&	-0.44$^{+0.05}_{-0.05}$	&	1.19$\pm{0.09}$ & 2.99$\pm$0.25 & 6.58$\pm$0.14$_F$	\\
MCG+04-22-042		&	0.033	&	1	&	S0		&	10.76	&	0.01$^{+0.03}_{-0.03}$	&	1.73$\pm{0.06}$ & 4.27$\pm$0.17 & 9.29$\pm$0.12$_F$ \\
MCG+08-11-011		&	0.020        &	1	&	SB0		&	11.20		&	0.59$^{+0.04}_{-0.04}$	&	15.85$\pm{0.14}$ & 44.71$\pm$5.06 & 109.00$\pm$5.06	\\
NGC 2110                      	&	0.007	&	2	&	SAB0	&	11.10		&	0.28$^{+0.03}_{-0.04}$	&	66.10$\pm{0.15}$ & 186.47$\pm$21.05 & 454.58$\pm$21.05 	\\
NGC 3516                      	&	0.009	&	1	&	SB		&	10.83	&	-0.15$^{+0.04}_{-0.09}$	&	5.28$\pm{0.14}$	& 14.90$\pm$1.72 & 36.3$\pm$1.72 \\
NGC 5548                      	&	0.017	&	1	&	SA0		&	10.62	&	0.23$^{+0.07}_{-0.07}$	&	4.52$\pm{0.11}$ & 11.23$\pm$0.31 & 24.42$\pm$0.14$_F$ \\
NGC 5728				&	0.010		&	2	&	SABa	&	10.58	&	0.32$^{+0.03}_{-0.06}$	&	7.88$\pm{0.06}$ & 22.23$\pm$2.52 & 54.19$\pm$2.52 	\\
UGC 11185 NED02           &	0.041	&	2	&	Irr/Int. 	&	10.72	&	0.54$^{+0.10}_{-0.16}$	&	8.17$\pm{0.06}$ & 23.05$\pm$2.61 & 56.18$\pm$2.61 \\
 \enddata
 
\tablecomments{(1) Object name, (2) optical spectroscopic redshift, (3) optical Seyfert type, (4) host galaxy morphology, typically taken from the NASA/IPAC Extragalactic Database (NED), (5) galaxy total stellar mass calculated using Equation~\ref{eq:mstar} [uncertainties in this estimate typically range from 0.05-0.15~dex according to \citet{Zibetti2009}], (6) star formation rate calculated using Equation~\ref{eq:sfr}, (7) total observed flux density at 22~GHz including both nuclear and extended components, (8) the flux density and errors at 5~GHz calculated either by interpolation, for those objects with a FIRST detection, and extrapolation for those objects outside the FIRST footprint (Section~\ref{sec:min_pressure}); and (9) the flux density and errors at 1.4~GHz as measured by FIRST, in the 3 cases bearing the subscript ``F", and as extrapolated in the remaining objects.}

 \end{deluxetable*}

\subsection{\emph{Herschel} Observations}
\label{sec:herschel}
The far-infrared \emph{Herschel}-observed sample consists of 313 BAT AGN. A description of the observations can be found in \citet{Melendez2014} and \citet{Shimizu2015}. The sample has both PACS and SPIRE imaging from \citet{Mushotzky2014}, providing sensitive 5-band photometry (70, 160, 250, 350, and 500 $\mu$m), extracted using manually-inspected circular and elliptical apertures to encompass the entirety of the far-IR flux. \citet{Shimizu2017} used analytical modeling to decompose the far-IR spectral energy distributions into contributions from the AGN and the host galaxy star formation, enabling an AGN-subtracted estimate of the star formation rate (Section~\ref{sec:mstarsfr}). 

\subsection{The BAT AGN Spectroscopic Survey}
\label{sec:bass}
All 11 of the objects presented in this paper have optical spectra from the BASS survey\footnote{www.bass-survey.com}, a large multi-wavelength effort to obtain optical spectra and other auxiliary data across the spectrum for the \emph{Swift}-BAT AGN. Eight have black hole mass estimates obtained either from the broad H$\beta$~emission line widths (for Type 1 AGN) or from the stellar velocity dispersion for narrow line AGN (Type 2 AGN). For NGC~3516 and NGC~5548, there are well-constrained masses from reverberation mapping studies \citep{Denney2010,Pancoast2014}. Only 2MASX~J0423+0408 lacks a black hole mass estimate as it does not have broad lines or a confident estimate of the velocity dispersion. This work also makes use of the stellar velocity dispersion measurements themselves (Section~\ref{sec:vcloud}), as well as in-progress BASS measurements from observations of CO (Section~\ref{sec:gravpot}). 

All of the objects in our sample have been observed in CO by an ongoing portion of the BASS multiwavelength follow up program \citep{Rosario2018,Lamperti2020}, and two works currently in progress (Koss et al. 2020, submitted, and Shimizu et al. in preparation). These observations occurred across three observatories: the James Clerk Maxwell Telescope (JCMT) on Mauna Kea, the 30m~millimeter telescope operated by the Institut de Radioastronomie Millim\'{e}trique (IRAM) in Spain, and the Atacama Pathfinder Experiment (APEX) in Chile.  Six targets have CO~1-0 fluxes or upper limits measured directly by IRAM (2MASX~J0423+0408, 2MASX~J1546+6929, MCG+04-22-042, MCG+08-11-011, NCG~5548, UGC~11185). Four targets have CO~2-1 fluxes measured by JCMT (MCG-01-40-001, NGC~2110, NGC~3516, NGC~5728) and one has a CO~2-1 flux measurement from APEX (MCG-02-08-014). 

\section{The Main Sequence of Star Formation}
\label{sec:mstarsfr}
\subsection{Calculation of Stellar Mass and SFR}
\label{sec:mass_sfr}

\citet{Shimizu2015} compared the BAT AGN sample to the star formation main sequence as calculated from the \emph{Herschel} Reference Survey \citep[HRS;][]{Boselli2010}, a complete sample of $\sim$300 inactive galaxies with a wide variety of morphological types and that has photometry matching that obtained for the BAT sample, and the \emph{Herschel} Stripe 82 Survey \citep[HerS; ][]{Viero2014}, which better matches the BAT AGN sample in stellar mass. 

The stellar masses ($M_*$) for most of our host galaxies have been calculated using the formula:

\begin{equation}
\mathrm{log}(M_*/L_i) = -0.963+1.032(g - i) \label{eq:mstar}
\end{equation}

from \citet{Zibetti2009}. This is the same equation used to calculate the main sequence from the HRS by \citet{Cortese2012}. The $g-i$ color was calculated using the AGN-subtracted PSF photometry from \citet{Koss2011}.

For two of our galaxies, the $g-i$~color is known to be contaminated by strong line emission, or the AGN subtraction is uncertain. For these two, 2MASX~J0423+0408 and 2MASX~J1546+6929, we use an alternative method described by \citet{Powell2018} (see their Section 2.1). Briefly, the full infrared SED was constructed from archival 2MASS and WISE photometry, corrected for Galactic reddening. The IR SED was then decomposed into an AGN component and a stellar component, which is fit by the stellar population synthesis models from \citet{Blanton2007}. We note that even though the small sample size could have allowed us to perform a SED-based IR estimate for all  soources, to be homogeneous with the HRS we preferred to use Equation~1 based on $g-i$ colors, and use the more detailed IR SED only for a couple of targets for which it was clear that the colors were contaminated from the AGN. Stellar masses are given in Table~\ref{t:tab1}. 

In order to determine the star formation rates (SFRs) from the infrared luminosity, it is first necessary to decompose the infrared spectral energy distribution (SED) into separate components arising from star formation and from the AGN. This was done for 313 targets from the \emph{Swift}-BAT survey by \citet{Shimizu2017}, using the \emph{Herschel} photometry described in Section~\ref{sec:herschel}. Here, we briefly summarize the fitting procedure. The contribution of star formation in a normal galaxy is frequently modeled as a modified blackbody that includes a frequency-dependent opacity, assumed to be a power law in the optically thin limit, and including the dust mass, spectral emissivity, and dust temperature. To account for the AGN, the model used by \citet{Casey2012} to study ULIRGs is adopted, in which the hot dust population is represented by a power law with an exponential cutoff. Unlike in \citet{Casey2012}, \citet{Shimizu2017} allows the power law normalization and the turnover frequency to vary freely, unconnected to the normalization of the modified blackbody component. This modification allows for the fact that AGN SEDs in the FIR are less continuous than those of ULIRGs: different heating processes are at work in the dust of AGN, while ULIRGs are powered by star formation throughout. The fitting itself utilizes a Bayesian framework with Markov chain Monte Carlo (MCMC) methods. Detailed descriptions, equations, and sample fits can be found in \citet{Shimizu2017}.

Once the star formation contribution to the IR emission is isolated, the SFR is calculated using the following relation from \citet{Murphy2011}:

\begin{equation}
\mathrm{SFR_{IR}}[M_\odot~\mathrm{yr}^{-1}] = \frac{L_{\mathrm{IR}} \mathrm{[erg~s^{-1}]}}{2.57\times10^{43}}. \label{eq:sfr}
\end{equation}

\noindent The resulting SFRs are given in Table~\ref{t:tab1}.

The calculations above were possible for all 11 of our KSR sources, and for 75 out of the 100 sources in the radio survey; see \citet{Shimizu2015} for details.


\begin{figure*}
\begin{tabular}{cc}
  \includegraphics[width=0.5\textwidth]{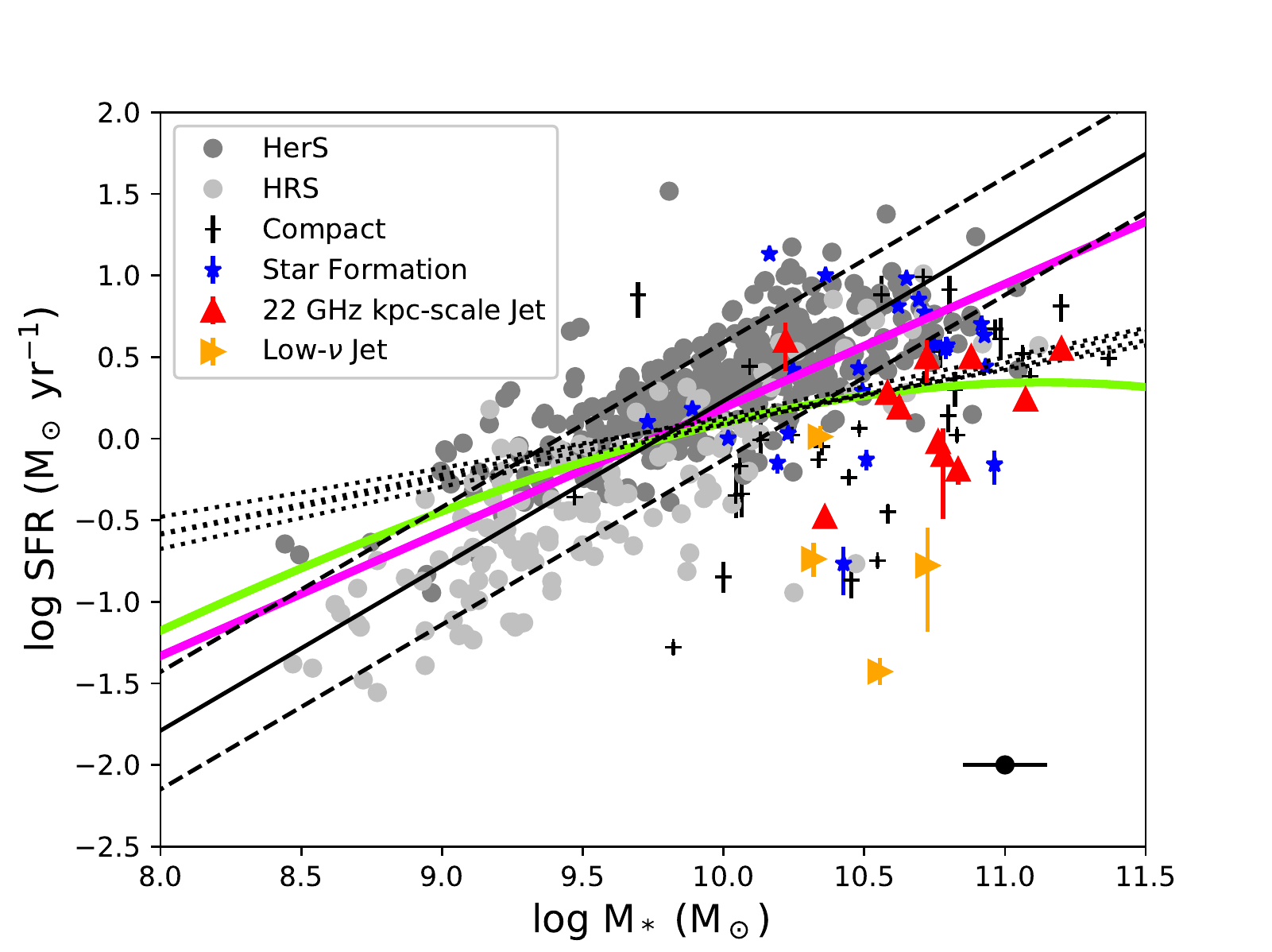}   & \includegraphics[width=0.51\textwidth]{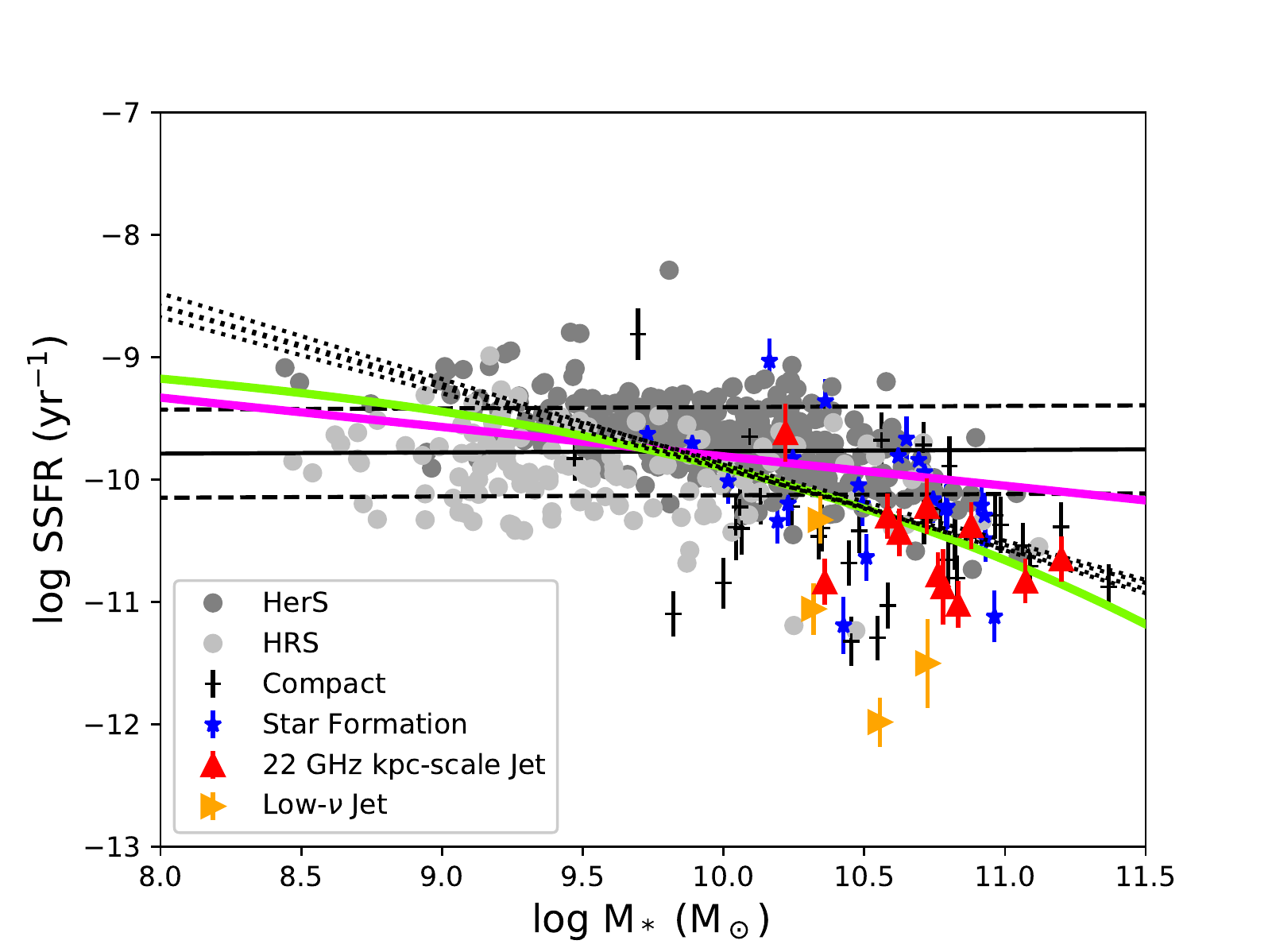} \\

\end{tabular}
\caption{\emph{Left}: The relation between stellar mass and star formation rate for the \emph{Herschel} Stripe 82 sample (dark grey circles),  the \emph{Herschel} reference survey (light grey circles) and BAT AGN (symbols denoting radio morphology as shown in the legend; see Section~\ref{sec:radio_obs}. The solid line represents the best fit to the star formation main sequence derived in \citet{Shimizu2015} from the HerS and HRS samples, and the dashed lines are the 1$\sigma$ scatter (0.36 dex). For context, we also show the main sequences as derived by \citet{Renzini2015} (pink), \citet{Saintonge2016} (green), and the four relations from different SFR-measurement methods from \citet{Popesso2019a} (black dotted); see Section~\ref{sec:discussion} for discussion.} \emph{Right}: The same, but showing the relation for the specific star formation rate (SSFR = SFR / M$_*$). The black point in bottom corner of each plot indicates the 0.15~dex error in the typical M$_*$ derived from $(g-i)$ color from \citet{Zibetti2009}; this error is used to calculate the errors in the SSFRs shown in the right panel. Colored curves are the same as in the left panel.
\label{fig:sfms}
\end{figure*}


\subsection{Suppressed Star Formation in AGN with Kiloparsec-Scale Radio Structures}
\label{sec:sfms}
In Figure~\ref{fig:sfms}, we plot the BAT AGN versus the star-forming main sequence, now including the radio morphological information from the survey. Errors in the SFR arise primarily from the uncertainties in the calculation of the SF-related component of the infrared luminosity as discussed in the previous section. Regarding errors in the stellar mass estimates, \citet{Zibetti2009} state that the rms of the empirical $M_* / (g-i)$ relation (Equation~1) ranges from 0.05-0.15~dex, equivalent to the effect of an error of 0.1~magnitudes in the color. We therefore make the conservative estimate of 0.15~dex on the errors of the stellar masses in the plots, and use this same assumption to calculate the errors in the specific star formation rate shown in Figure~\ref{fig:sfms}. With one exception, all sources in our KSR sample are at least 1$\sigma$~ below the main sequence, while objects with nuclear star formation radio morphologies reside on the main sequence. That exception is 2MASX~J04234080+0408017,  the object with the largest radio structure in the sample, spanning approximately 20 kpc and far exceeding the host size (see Figure~\ref{fig:hosts}). We discuss this object in more detail in Section~\ref{sec:discussion}. 

It is known that star formation in Seyferts can be suppressed by disk truncation due to a cluster environment; for example, more than half of the galaxies in the Virgo cluster have truncated H$\alpha$ disks \citep{Koopmann2004}. We have therefore investigated whether or not the observed suppression is due to the objects' environment, by both visual inspection and by checking the 2MASS group catalog by \citet{Tully2015}. Only one galaxy, NGC~5548, belongs to a cluster, AWM~3. This is a ``poor" cluster identified by \citet{Beers1995}. Of the other 10 galaxies, four are found in groups by \citet{Tully2015}, but the groups consist of only 3-5 galaxies. It is unlikely that environmental stripping is playing a significant role in star formation suppression for this sample.

The shape of the star forming main sequence has been much debated over the past decade, especially in the high-$M*$ regime (log$M_* / M_\odot > 10.5$). Depending on selection methods, and especially if star forming galaxies are pre-selected by color to form the main sequence, the shape of the relation can become flatter for high mass galaxies. We have followed \citet{Shimizu2015}, defining the main sequence using the \emph{Herschel} reference surveys, since these surveys have stellar masses and star formation rates calculated in a self-consistent manner with the BAT sample. Our relation is similar to the objective description for the main sequence as calculated from a sample of $\sim240,000$ galaxies with $0.02 < z < 0.085$ from SDSS DR7 by \citet{Renzini2015}, defining the relation as the ridgeline of the three-dimensional distribution of SFR, $M_\odot$, and number. They do not find any indication of a flattening at high masses, and claim that such a flattening is often an artefact of selection effects. We do note, however, that \citet{Saintonge2016} used massive, local galaxies in SDSS DR7 (log$M_* / M_\odot > 10$, $0.025 < z < 0.05$) and fit a cubic polynomial to the ridgeline, finding a relation that bends downward. \citet{Popesso2019b} explored the high-mass end of the main sequence using a larger sample from SDSS DR7 extending to much higher redshifts and four different methods of measuring the SFR, and found linear fits, but at slightly shallower slopes than \citet{Renzini2015}. For context, we provide all of these relations in Figure~\ref{fig:sfms}.

\section{Feedback Energetics: A Jet Interpretation}
\label{sec:energetics}

In order to determine whether the observed jets are capable of supplying sufficient energy to heat or evacuate the gas, we must first estimate the total jet power from the observed luminosity. Unfortunately, there is currently no theoretically supported way to do this for small, nonrelativistic kiloparsec-scale outflows. To proceed as conservatively as possible, we select three different methods by which to estimate the jet power, and then compare each estimate to two criteria for successful negative feedback: the ability to remove the molecular gas from the galaxy gravitational potential, and the kinetic energy transfer to gas clouds leading to cloud velocities $v_\mathrm{cloud}$ exceeding the stellar velocity dispersion $\sigma_*$.

\subsection{Jet Power Estimates}
\subsubsection{Minimum Equipartition Pressure}
\label{sec:min_pressure}

First, we estimate the jet power by assuming that the observed radio luminosity is due to incoherent synchrotron emission from relativistic electrons radiating in the magnetic field of the jet. If we assume equipartition between the magnetic energy density and the energy density of the electrons, we can obtain an estimate of the minimum pressure in the jet lobes. In practical units, the expression for minimum pressure is $p_\mathrm{min} \sim 10^{-12} (T_5 / l_\mathrm{kpc})^{4/7}$~dyn~cm$^{-2}$, where $T_5$~ is the brightness temperature at 5~GHz and $l_\mathrm{kpc}$~ is the extent of the jet along the line-of-sight direction in kiloparsecs \citep{Begelman1984}. 

To estimate the 5~GHz brightness temperature, we first estimate the 5~GHz flux density by interpolating between the total 22~GHz flux density and the 1.4~GHz archival flux density from the FIRST survey \citep{Becker1995} for the 3 objects in our KSR sample within the FIRST footprint: MCG-02-08-014 ($S_{\nu,\mathrm{1.4GHz}} = 6.58$~mJy), MCG+04-22-042 ($S_{\nu,\mathrm{1.4GHz}} = 9.29$~mJy), and NGC~5548 ($S_{\nu,\mathrm{1.4GHz}} = 24.42$~mJy). For the 8 objects without FIRST coverage, we extrapolate to 5~GHz assuming a spectral index of $\alpha=-0.7$, where the flux density $S_\nu \sim \nu^{\alpha}$. Although this is a standard scaling, because the origin of the radio emission (i.e., whether a steep spectrum is a reasonable assumption) is relevant to this work, we calculate the spectral indices for the three objects with FIRST measurements as a consistency check on the validity of this assumption. We obtain $\alpha = -0.62\pm0.14, -0.61\pm0.11$, and $-0.61\pm0.15$ for MCG-02-08-014, MCG+04-22-042, and NGC~5548 respectively. A ``steep" spectral slope of $\alpha\sim-0.7$ and resolved, biconical structure on arcsecond scales are indications of an outflow, non-jetted origin of radio emission in radio-quiet AGN, as nicely illustrated in the recent overview by \citet{Panessa2019}.

With the estimated 5~GHz fluxes in hand, we then calculate the brightness temperature:

\begin{equation}
T_b = \frac{\lambda^2 S_\nu}{2k\Omega}
\end{equation}

where $S_\nu$~is the flux density, and $S_\nu / \Omega = I$, the intensity. If the intensity is  measured in Jy per sterradian, $I = 2760 \times T_b / \lambda^2$, from which we can estimate the brightness temperature at 5 GHz. 

The next requisite quantity is the line-of-sight extent of the jet, which requires an estimate of the inclination of the jet axis. It is necessary in this case to resort to a rather crude estimation, which can be obtained from the optical Seyfert type. Although the jet can be oriented at any angle with respect to the host galaxy, it is always emitted perpendicular to the accretion disk. In Type~2 AGN the dusty torus is thought to block the broad-line emission due to relatively high inclinations to the line-of-sight. In more face-on orientations, broad emission lines are clearly visible in the optical spectrum (Type 1). For this order-of-magnitude calculation, we simply posit an inclination $i = 30^{\circ}$ for objects with Type 1 optical spectra, and $i=70^{\circ}$ for Type 2. We note that such an assumption neglects additional reasons why objects may lack broad lines, including severe dust extinction in the host galaxy or alternative accretion flow geometries. We then calculate the line-of-sight extent of the jet from the apparent length of the jet in the plane of the sky as $l_\mathrm{kpc,LOS} = l_\mathrm{kpc,app} / \tan i$.

We insert the values for $T_b$~and $l_\mathrm{kpc,LOS}$~into the formula for $p_\mathrm{min}$~above. To obtain a rough estimate the energy in the jet, we then multiply this pressure by the jet volume. The volume is obtained by approximating the jet as a cylinder, the ``height" of which is the true jet extent, $l = (l_\mathrm{LOS}^2 + l_\mathrm{app}^2)^{1/2}$, and the radius of which is estimated by measuring the distance in the images from the brightest point in the jet lobe to the edge of the lobe, taken as the 5\% radio contour. The energies estimated by this first method are given in Table~\ref{t:energy} as $E_1$. 

\subsubsection{Energy in Relativistic Electrons}
\label{sec:electron_energy}
Next, we estimate the jet energy following the arguments in \citet{Leahy1991},  referred to as L91 in this section. Assuming a constant magnetic field and a power law spectral shape, one can integrate over the source to obtain an estimate of the total energy density in relativistic electrons $U_\mathrm{re}$ that does not depend on geometry or line-of-sight considerations (L91 Section~3.2.7). The energy flux producing the relativistic electron energy is expressed as $K_\mathrm{re} = (4/3) U_\mathrm{re} / \tau$, where $\tau$~is the radiative lifetime of the electrons estimated based on equipartition arguments and spectral ageing considerations; for full details, see L91~Sections 3.2.5 and 3.2.6 for details. The expression for $K_\mathrm{re}$, found in L91~Section~3.4.4.3, then becomes:

\begin{equation}
K_\mathrm{re} = (2^8 3^{-9/2} \pi^2) L_{\nu0} \nu_0^\alpha \mathcal{I} _1 \nu_{T_{lobe}}^{1/2} [1+(B_\mathrm{MB} / B)^2] \label{eq:kre}
\end{equation}

\noindent where $L_{\nu0}$~is the luminosity at some reference frequency $\nu_0$, in this case 22~GHz; $\nu_{T_{lobe}}$ is the frequency at which the lobe spectrum steepens to a high-frequency regime dominated by radiation losses, assumed here following convention to be 100~GHz; and $B_\mathrm{MB}$ is the equivalent magnetic field to the cosmic microwave background radiation, $B_\mathrm{MB} = 0.318(1+z)^2$~nT, which is on the order of $\sim0.3$~nT for redshifts considered here. In this expression, $\mathcal{I}_1$~is a weighted integral over the spectrum that depends on $\nu_\mathrm{upper}$~and $\nu_\mathrm{lower}$, the frequency cutoffs of the power-law portion of the spectrum, and on the spectral index $\alpha$. Typically,  $\nu_\mathrm{lower}$~is set to 10~MHz and $\nu_\mathrm{upper}$ to 100~GHz, and we continue to assume $\alpha=-0.7$. Discrepancies in the values of $\nu_\mathrm{upper}$~and $\nu_\mathrm{lower}$ do not affect $\mathcal{I}_1$ by much more than a factor of 2; see Equation~3.15 in L91. 

The main uncertainty in Equation~\ref{eq:kre} is $B$. In L91, the convention was $B=0.1$~nT based on observations of the Coma cluster. This is many orders of magnitude smaller than the magnetic field measured from VLBI observations of parsec-scale blazar jets very near the black hole, which tend to be a few thousand nT \citep{OSullivan2009}. If $B\gg B_\mathrm{MB}$, the final bracketed term in Equation~\ref{eq:kre} is approximately unity. If we assume the L91 value of 0.1 nT, the bracketed term is approximately a factor of 10. We assumed the latter value in order to conservatively estimate a jet power to be compared later with feedback requirements.

After inserting the observed 22~GHz luminosities for each target into these formulae, we obtain the resulting energy estimates $E_2$, given in Table~\ref{t:energy}. 

Because the radio lobes presumably expand as they age, the total energy will be larger. As pointed out by L91, $K_\mathrm{tot} = 2 ( 1+k)~K_\mathrm{re}$ after following equipartition arguments, where $(1+k)$~ is the ratio of energy density from all particles to that from relativistic electrons. The population of other particles is not well constrained. Therefore, this jet energy is strictly a lower limit.

\subsubsection{Energy Required by X-ray Cavities}
\label{sec:e3_cavity}

Our final method of estimating the energy within the jet relies upon an empirical correlation between the radio luminosity at 5~GHz and the jet power required to create and fill observed X-ray cavities or bubbles. \citet{Merloni2007} estimated the work necessary to inflate hot, X-ray emitting cavities in AGN host galaxies and clusters, and found a correlation between the required kinetic power and observed radio emission: log~$L_\mathrm{kin} = (0.81\pm{0.11})~\mathrm{log} L_\mathrm{R}~+ ~11.9^{+4.1}_{-4.4}$, which agrees well with synchrotron jet models from theory. \citet{Mezcua2014} used this relation to estimate the jet power in radio structures spanning $0.1 - 3.8$~kpc in low-luminosity AGN very similar to our sample. 

We obtain the 5~GHz luminosities by the same method described in Section~\ref{sec:min_pressure}. After arriving at the 5~GHz luminosities, we scale them to the jet power estimate using the relation from \citet{Merloni2007} above. We call these $E_3$, and provide them in Table~\ref{t:energy}.

\subsection{Energy Required to Suppress Star Formation}

We now compare our three energy estimates based on the jet interpretation to two criteria for effective suppression of star formation by AGN feedback: the energy required to remove molecular gas from the host galaxy, and the energy required to accelerate gas clouds to a velocity higher than the stellar velocity dispersion.

It is not appropriate to compare the entire estimated jet energy to such criteria, however. Most of our host galaxies are spirals (see Section~\ref{sec:gravpot}). Although the energy deposited into an inhomogenous ISM by jets can be quite efficient in elliptical galaxies \citep{Wagner2011}, other simulations show that even in elliptical galaxies the fraction of the AGN power deposited into the ISM is only a few percent \citep{Cielo2014}; see also the review by \citet{Harrison2018}. It is natural to assume that in a disk galaxy, far less of the ISM will be impacted by a jet that is most likely to propagate perpendicular to it. However, inclined jets that are directed at least partially into the disk plane can cause substantial disruption to the ISM and affect star formation \citep{Cielo2018}. As the premise of this paper is to determine the most likely mechanism by which such structures impact the ISM, we adopt a generous blanket efficiency factor of jet power to ISM coupling of 10\%, higher than but on order of the few-percent assumptions in the papers mentioned above.

\subsubsection{Gravitational Potential of Hosts}
\label{sec:gravpot}
The first ingredient in the gravitational potential is the total mass of the galaxy. We begin by using the stellar mass calculated earlier (Section~\ref{sec:mstarsfr}), and recognize that there is likely to be an additional large contribution to the mass from dark matter. As we do not have rotation curves for these galaxies, we must assume a mass-to-light ratio. 

Mass-to-light ratios depend upon the galaxy Hubble type. We have obtained the morphology for each host galaxy from the NASA/IPAC Extragalactic Database (NED)\footnote{http://ned.ipac.caltech.edu}, or estimated it ourselves based on the images in Figure~\ref{fig:hosts} in the case that NED classifies it as an AGN only; the morphologies are given in Table~\ref{t:tab1}. Ten of the 11 galaxies are clearly spirals, most of which are relatively early-type. The exception is UGC~11185, a pair of merging galaxies hosting a dual obscured AGN \citep{Hainline2016}. From the review by  \citet{Faber1979}, the typical mass-to-light ratio of an early type spiral is $\sim$4. We therefore multiply the stellar masses by 4, to obtain an estimate of the total galaxy mass. 

Next, we require the approximate size of the galaxy. We simply measure the angular sizes of the hosts as shown in Figure~\ref{fig:hosts} and convert them to physical distances assuming the standard cosmology described in the Introduction. The gravitational potential energy of the molecular gas is then just $U_\mathrm{grav} = (3/5)~GM_\mathrm{tot}M_\mathrm{gas}/R$, where $M_\mathrm{gas}$~is the molecular gas mass.

All 11 targets have CO flux measurements from ongoing work within the BASS collaboration (Section~\ref{sec:bass}). Nine targets have detections in CO, and two (2MASXJ~1546+6929 and MCG-02-08-014) have upper limits only. Five targets have flux measurements of the CO~2-1 transition, rather than CO~1-0. Since it is usual to measure the gas mass from the 1-0 transition, we must convert these measurements into CO~1-0 fluxes via a conversion factor. The ratio of CO~2-1 / CO~1-0 fluxes ranges from $\sim0.7$~to $\sim1.1$~in galaxy samples comparable to ours \citep{Braine1992,Sandstrom2013}; we therefore adopt a value of 0.85 for these calculations. 

We next convert the CO 1-0 fluxes to luminosities

\begin{equation}
L_\mathrm{CO}' = 3.25 \times 10^7 ~S_\mathrm{CO} ~\nu_\mathrm{obs}^{-2} ~D_L^2 ~(1+z)^{-3}~\mathrm{K~km~s}^{-1}~\mathrm{pc}^2
\end{equation}

where $S_\mathrm{CO}$ is the CO~1-0 flux in units of Jy~km~s$^{-1}$, $\nu_\mathrm{obs} = 115$~GHz (the frequency of the CO~1-0 transition), and $D_L$~is the luminosity distance in megaparsecs. To convert this to a mass of molecular hydrogen, one requires a CO-to-H$_2$~conversion factor $\alpha_\mathrm{CO}$. For the Milky Way, $\alpha_\mathrm{CO} = 4.4$~M$_\odot$pc${^{-2}}$/(K km s$^{-2}$) \citep{Bolatto2013}. We multiply our $L_\mathrm{CO}'$~values by this factor, thereby arriving at the molecular gas masses given in Table~\ref{t:energy}, which can be inserted into the formula for $U_g$~to obtain the energy required to eject it from the host galaxy, also given in Table~\ref{t:energy}. 

We note that there is wide variation in the value of $\alpha_\mathrm{CO}$ across galaxies, with a few AGN showing somewhat lower values than the Milky Way (1.2 - 3), and many star-forming galaxies showing comparable values to the Milky Way, or a bit higher \citep{Sandstrom2013}. There are no galaxies in that study with values of $\alpha_\mathrm{CO}$ different from the Milky Way by more than a factor of a few, and certainly not more than an order of magnitude, so the results on our simplistic calculation is likely to be minimal.

\subsubsection{Velocity Imparted to Clouds}
\label{sec:vcloud}
An alternative feedback criterion is whether the jet or outflow can impart a velocity to gas clouds, $v_\mathrm{cloud}$, that exceeds the velocity dispersion, $\sigma_*$, of the host galaxy \citep{Silk1998}. As pointed out by \citet{Wagner2011}, this does not necessarily mean these clouds will be ejected, but they would be effectively dispersed. 

We will use the Milky Way as the proxy by which we estimate the molecular cloud properties for our sample, as they do appear to be spirals, and we know more about the Milky Way's distribution of molecular gas than any other galaxy. 

Molecular cloud masses range from a few solar masses to as much as $\sim10^7$~M$_\odot$, where this cutoff mass varies from $10^4 - 10^6$~M$_\odot$ in the Milky Way and other Local Group galaxies \citep{Rosolowsky2005}. The vast majority of the mass is contained in the largest molecular clouds.

In the Milky Way, 90\% of the molecular gas is contained in clouds with masses greater than $\sim5\times10^5$~M$_\odot$, numbering approximately 5000 clouds based on a detailed accounting of the CO emission in the Milky Way \citep{Miville2017}. After this point, the cumulative mass distribution drops off as a steep power law towards the cutoff mass \citep[e.g., ][]{Solomon1987,Rosolowsky2005}. 

To determine if 5000 clouds is a reasonable amount for the galaxies in our sample, we use the molecular gas masses calculated in the previous section and given in Table~\ref{t:energy}. We first take 90\% of these masses (to account for the 10\% likely contained in the small clouds), and divide them by $5\times10^5$, a representative mass for each cloud as described in the preceding paragraph. The mean value for the number of clouds of that mass for our sample is 4786, quite close to the value reported for the Milky Way.

We therefore proceed to use these parameters to estimate the velocity that the outflows can impart to the clouds. Dividing our energy estimates by 5000, we obtain the energy imparted to a single cloud, $E_\mathrm{cloud}$. If we make the conservative assumption that all of this energy is converted into kinetic motion of the entire cloud, then the resulting $v_\mathrm{cloud} = (2E_\mathrm{cloud}  / M_\mathrm{cloud})^{1/2}$.  Note that these estimates are the maximum possible energy imparted, since in reality some fraction of the energy would be converted into heat and some energy would miss interacting with the gas entirely in an inhomogeneous ISM, which is almost certainly the case.

We must then compare these imparted velocities to $\sigma_*$. The stellar velocity dispersions $\sigma_*$~have been measured from optical spectra for 8 of our targets in the BASS survey, given in Table~\ref{t:energy}. We do not have an estimate of $\sigma_*$~for 3 targets because the bright, unobscured AGN made the stellar absorption line properties difficult to measure.  

\vspace{\baselineskip}
The results of all of these tests are shown in Figure~\ref{fig:energetics}, with the above energy estimates shown in red, and required thresholds of the host gravitational potential energy and the stellar velocity dispersions, as calculated in the immediately preceding sections, shown as horizontal dashed lines. To demonstrate the uncertainties, we have included error bars, which are usually  small compared to the measured quantities, propagated from the measurement errors in the 22~GHz luminosity through any necessary extrapolation or interpolation. For $E_1$, the propagated errors also include a 15\% measurement error on the apparent size of the jet, based off of multiple trials. We have also included uncertainty bands in the figure, to illustrate the range of possible values if other assumptions than ours are made: for $E_1$, the band encompasses an inclination range of $5^{\circ} < i < 40^{\circ}$ for Type~1 AGN and $50^{\circ} < i < 85^{\circ}$ for Type~2 AGN (Section~\ref{sec:min_pressure}); for $E_2$ the band demonstrates the result of choosing $B = 0.1$nT instead of $10^3$nT for the magnetic field strength in Equation~4 (Section~\ref{sec:electron_energy}); and for $E_3$ the band shows the 0.47 dex scatter in the \citet{Merloni2007} relation between radio luminosity and kinetic power (Section~\ref{sec:e3_cavity}).


\begin{deluxetable*}{lcccccccc}
 \tablewidth{0pt}
 \tablecaption{BAT AGN with Kiloparsec-Scale Radio Structures: Energetics Estimates\label{t:energy}}
 \tablehead{
 \colhead{Name} &
   \colhead{log $p_\mathrm{min}^{-9}$} &
  \colhead{log $ E_1$} &
  \colhead{log $ E_2$ } &
  \colhead{log $ E_3$} &
  \colhead{log $ E_\mathrm{wind}$} &
  \colhead{log $M_\mathrm{H_2}$} &
  \colhead{log $U_\mathrm{grav}$} &
  \colhead{$\sigma_*$} \\
  \colhead{} &
   \colhead{\footnotesize $10^{-9}$dyn~cm$^{-2}$} & 
 \colhead{\footnotesize erg} &
 \colhead{\footnotesize erg} &
 \colhead{\footnotesize erg} &
 \colhead{\footnotesize erg} &
 \colhead{\footnotesize $M_{\odot}$} &
 \colhead{\footnotesize erg} &
\colhead{\footnotesize km s$^{-1}$} \\
 \colhead{\tiny (1)} &
 \colhead{\tiny (2) } &
 \colhead{ \tiny (3) } &
 \colhead{ \tiny (4) } &
 \colhead{ \tiny (5) } &
 \colhead{ \tiny (6) } &
 \colhead{ \tiny (7)} &
 \colhead{\tiny(8)} &
 \colhead{\tiny (9)} \\
}

 \startdata 
2MASX J04234080+0408017       	&	0.105	&	57.5	&	54.2	&	57.7	&	58.4	&	9.25	&	55.8	&	-	\\
2MASX J15462424+6929102       	&	0.013	&	54.0	&	53.3	&	57.0	&	57.6	&	8.52	&	55.61*	&	288 14	\\
MCG-01-40-001	&	0.173	&	54.3	&	54.0	&	57.6	&	58.4	&	10.05	&	57.3	&	-	\\
MCG-02-08-014	&	0.006	&	53.9	&	52.5	&	56.3	&	56.7	&	9.00	&	55.67*	&	138 4	\\
MCG+04-22-042	&	0.001	&	54.8	&	53.3	&	56.9	&	57.4	&	9.24	&	56.3	&	180 36	\\
MCG+08-11-011	&	0.073	&	53.7	&	53.8	&	57.4	&	58.1	&	9.20	&	56.7	&	-	\\
NGC 2110                      	&	0.421	&	53.2	&	53.6	&	57.3	&	57.9	&	8.54	&	56.0	&	299 19	\\
NGC 3516                      	&	0.104	&	55.6	&	52.8	&	56.7	&	57.2	&	9.05	&	56.2	&	483 48	\\
NGC 5548                      	&	0.014	&	54.4	&	53.1	&	56.8	&	57.3	&	9.29	&	56.2	&	180 48	\\
NGC5728	&	0.075	&	53.2	&	52.5	&	56.4	&	56.9	&	9.71	&	56.6	&	160 11	\\
UGC 11185 NED02               	&	0.073	&	54.5	&	54.1	&	57.7	&	58.4	&	9.47	&	56.5	&	164 17	\\
 \enddata
 
 \tablecomments{ (1) Object name, (2) minimum pressure in the jet from equipartition arguments from Section~\ref{sec:min_pressure}, (3) jet energy based on the minimum pressure from Section~\ref{sec:min_pressure}, (4) jet energy based on the kinetic energy in relativistic electrons from Section~\ref{sec:electron_energy}, (5) jet energy based on scaling relations from radio luminosity and X-ray cavities from Section~\ref{sec:e3_cavity}, (6) energy of a radiatively-driven AGN wind assuming a lifetime of $10^7$~yr as calculated in Section~\ref{sec:wind}, (7) molecular gas mass calculated from the CO luminosity as described in Section~\ref{sec:gravpot}, (8) energy required to eject the molecular gas mass from the galaxy calculated as described in Section~\ref{sec:gravpot}, (9) stellar velocity dispersion of the host galaxy measured from optical spectroscopy. Asterisks on quantities denote upper limits.}

 \end{deluxetable*}

\section{Feedback from an AGN-driven Wind}
\label{sec:wind}

Up until now, we have considered the observed radio morphologies as small, possibly immature or frustrated radio jets analogous to the larger, more powerful relativistic jets found in radio galaxies. However, there is another important interpretation of such radio structures. Instead of a jet, radio emission may arise as a secondary consequence of a radiatively-driven AGN outflow interacting with the ISM \citep{Zakamska2014}. The biconical nature of the outflow can result in a pseudo-jet structure in radio emission. 

In this case, the radio morphology is evidence of an underlying AGN wind: the actual source of energy that is potentially suppressing the host star formation. As found by \citet{Zakamska2014}, the efficiency of converting the wind kinetic energy into radio luminosity is $3.6\times 10^{-5}$, for radio emission from the FIRST survey at 1.4~GHz. This encapsulates similar information to the coupling coefficient used in the above estimates (i.e., only the gas that ``notices" or is affected by the wind emits radio emission). We can therefore estimate the energy available in the underlying wind by dividing the observed radio luminosity, either as measured by the FIRST survey or calculated by extrapolation, by this efficiency factor. As stated before, three of our targets have FIRST coverage and detections; for the remaining objects we assume a spectral index of $\alpha=-0.7$, see Section~\ref{sec:min_pressure}. We then use Equation~2 from \citet{Zakamska2014} to convert from the 1.4~GHz flux densities to luminosity, which includes the redshift $k$-correction.

Given the power, to calculate the energy available for injection into the environment we must assume a lifetime. One handy estimate is simply the typical quasar lifetime (note that this is distinct from the radiative lifetime of relativistic electrons used to estimate jet power in Secton~\ref{sec:electron_energy}). Observationally, the length of time that a quasar is ``on" is often found to be $\sim10^7-10^8$ years in models \citep{Gan2014}; this is consistent with many observations \citep{Martini2001,Marconi2004,Bird2008}. It is also important to distinguish between the lifetime of the optical AGN phase and that of the radio outflows. Some observations and simulations support optical lifetime estimates as short as $10^5$ years \citep{Schawinski2015, Yuan2018}, while the outflows may last longer than the lifetime of the quasar; see \citet{Zubovas2018} and references therein. We therefore decide upon a value of $10^7$ years to determine the energy supplied by the wind for this work. In order to check the validity of this estimate, we can make a rough calculation of the minimum lifetime using the same linear extents reported in Section~\ref{sec:min_pressure} and assuming an underlying wind velocity of 1000~km~s$^{-1}$. This velocity is supported by many observations of molecular and ionized AGN outflows \citep[e.g., ][]{Kraemer2000,Cicone2012} as well as the velocity of X-ray warm absorbers in Seyferts \citep[e.g., ][]{Mathur1995,Reynolds1997}. Because the energy is simply the product of the power and the lifetime, the reader can rescale the estimated supplied energy by an order of magnitude in either direction to arrive at the number for longer or shorter lifetimes.

We show the energy available in the wind responsible for the secondary radio emission for two different fiducial lifetimes in blue in Figure~\ref{fig:energetics}. As for the jet energy estimates, we provide uncertainty bands in the figure; in this case, they come from the uncertainty in applying the calibration between the outflow energy and the observed luminosity from \citet{Bell2003} used by \citet{Zakamska2014} in their analysis. This is approximately a factor of 2, or $\sim0.3$~dex.

\section{Discussion}
\label{sec:discussion}

In determining whether or not the observed radio structures are frustrated AGN jets or radio byproducts from AGN-driven outflows, accretion rate may play a role. The launching of traditional AGN jets tends to occur in low-accretion rate objects \citep[see the review by ][]{Blandford2019}. However, the distribution of Eddington ratios among our sample of 11 KSR objects is not different from the overall Eddington ratio distribution \citep{Smith2020a}. This supports the paradigm that these are radio byproducts of radiatively-driven AGN outflows.

\begin{figure*}
   \centering
    \includegraphics[width=\textwidth]{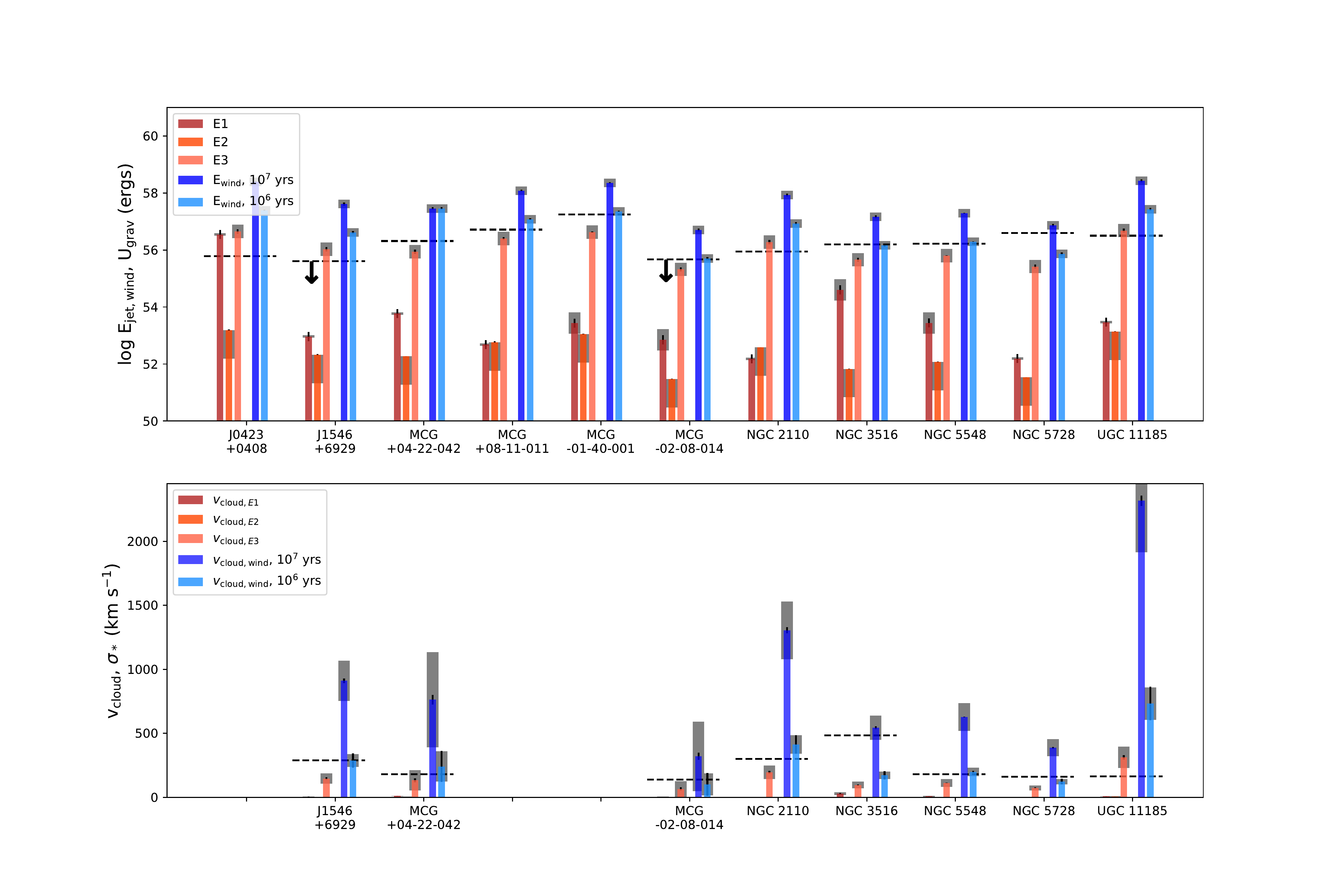}
    \caption{Comparison of energy estimates with various required quantities for successful star formation suppression via feedback. \emph{Top:} histogram of energies derived using each method $E_1$, $E_2$, and $E_3$ in shades of orange. Wind energies derived for lifetimes as denoted in the legend shown in blue. Gravitational potential of the galaxy and halo as calculated in Section~\ref{sec:gravpot}, and therefore the required threshold for effective feedback, are given for each object as dashed horizontal lines. 
    For the objects with upper limits for $L_\mathrm{CO}$, and therefore on the gas masses and potentials, black arrows are shown. Black error bars (typically very small) indicate the error in the quantity as propagated from the error in the measured or extrapolated/interpolated radio luminosity. Grey bands indicate the uncertainties in the quantities due to the assumptions made in the calculations; see Sections~\ref{sec:energetics} and \ref{sec:wind}  for details. \emph{Bottom:} estimates of the cloud velocity imparted by the jet or wind given each energy estimate. Colors are the same as in the upper panel. Horizontal dashed lines indicate the measured value of $\sigma_*$ from optical spectra; feedback is effective when $v_\mathrm{cloud} \geq \sigma_*$. Three objects are excluded from the bottom panel due to lack of measurements of $\sigma_*$; many values on the bottom panel are too low to see, being only a few or a few tens of km/s. Error bars and uncertainty bands are the same as in the top panel.}
    \label{fig:energetics}
\end{figure*}

In Figure~\ref{fig:energetics}, we plot histograms of the jet energies $E_1$, $E_2$, and $E_3$, as well as the wind energies for two lifetimes, and their implied  $v_\mathrm{cloud}$~ values. The values of the necessary thresholds, ($U_\mathrm{grav}$, $\sigma_*$) are given as horizontal dashed lines.   

Some caveats are needed before discussing whether or not the estimated energies could produce the suppression of star formation evident in Figure~\ref{fig:sfms}. First, we should consider how much of the cold gas reservoir must be affected. Most of the galaxies have SFRs that are approximately 10\% of normal for their stellar mass if the \emph{Herschel}-defined main sequence is adopted; star formation has not entirely ceased. Therefore, even if an energy estimate is not above the threshold, it could have a meaningful impact on star formation.  It is also important to note that the energy calculations based on equipartition arguments are expressions of the \emph{minimum} jet power. 

Next, we consider the shape of the main sequence. Numerous main sequences exist in the literature, with various slopes and shapes. As mentioned in Section~\ref{sec:mass_sfr}, some previous studies have found star forming main sequences that flatten at higher masses \citep[e.g., ][]{Saintonge2016,Popesso2019a}. Other studies have found a variety of constant slopes, with no indication of flattening \citep[e.g., ][]{Sargent2014,Renzini2015}. The shape depends strongly upon the selection criteria, and the normalization and shape may also evolve with redshift \citep{Whitaker2012,Speagle2014}; although other authors do not find an evolution of the high mass slope with redshift \citep{Popesso2019b}. In the spirit of consistency, we define the main sequence using the \emph{Herschel} reference surveys as in \citet{Shimizu2015}, since these samples have measurement methods of the SFR and stellar mass that are consistent with the methods used for the BAT AGN. The sample used to calculate the main sequence has also been redshift-limited to be consistent with the BAT sample ($0.02 < z < 0.08$). Still, we provide bending and shallower main sequence relations in Figure~\ref{fig:sfms}, so that readers familiar with these relations have context for our sample. The majority of our objects remain below the \citet{Renzini2015} main sequence, which is the closest-matching to ours in selection methods.

In the majority of objects (7/11), none of the energies calculated assuming the structure is a radio jet exceeded the thresholds for gas evacuation. In 2MASX~J0423+0408, two jet estimates exceed the gravitational threshold; this is mainly due to the very large physical size of the radio emission. In 2MASX~J1546+6929, one jet estimate exceeds the gravitational threshold. UGC~11185 is a marginal case in which one jet estimate exceeds the cloud velocity threshold, and the same estimate essentially meets the gravitational threshold.

By contrast, in every target, the energy estimate from assuming the KSR indicates an underlying wind meets or exceeds the gravitational evacuation threshold, and in all but one case (NGC~3516), exceeds the cloud velocity threshold, often by a wide margin.

Evidence for an AGN-driven disk wind has been found for few of our targets based on X-ray observations, supporting the existence of the hypothesized radiative outflow: \citet{Reynolds1997} and \citet{Turner2008} both find evidence for a rapidly outflowing warm absorber in the X-ray spectrum of NGC~3516, and \citet{Mathur1995} find such a signature in NGC~5548. 

We can conclude from these simple calculations that the kinetic influence of the radiative outflow indicated by the observed radio emission is significant, that the observed suppression is likely to be due to a radiative outflow, and that the observed radio structures are the signatures of that outflow's ISM interaction. This interpretation may also explain why the radio structures are in many cases much smaller than the host galaxy, which is nevertheless suffering a global SFR suppression. 

There is evidence from recent detailed studies of AGN hosts versus inactive galaxies with optical IFU data that feedback from such outflows is an important and so-far underestimated player in the evolution of galaxies \citep{Wylezalek2020}, and theoretical indications that such feedback is required in low-mass galaxies to conform cutting-edge cosmological simulations to observations \citep{Davies2019}.

Although the calculations in this study are simplified and assume a straightforward interaction of the jet or outflow with the gas, detailed theoretical modeling has recently proposed a highly effective feedback mechanism that requires less radiative energy than is directly calculated from radio luminosities. \citet{Hopkins2010} demonstrate a ``two-phase" model in which a small nuclear jet or other energy deposition method drives an outflow in the hot interstellar medium, which then travels outward to kiloparsec scales. As this happens, the cold molecular clouds it encounters expand perpendicular to the outflow and are effectively shredded, suppressing star formation even at large galactic radii. Quantitatively, this two-phase model reduces the required energy for effective feedback from direct entrainment or heating by factor of at least the mass fraction of the hot interstellar medium, which is about $f_\mathrm{hot} \sim 0.1$. Such a consideration would bring the jet-based energy estimates into feasibility for most of our targets, but would also increase the efficacy of the radiative outflow scenario. 

This is not to say that no kiloparsec scale radio structures in Seyferts are genuine radio jets, or that such jets would not impact the star formation of their hosts in a meaningful way.  \citet{Jarvis2019} found strong indications of kiloparsec-scale jet interaction with the warm ionized gas of the ISM in optical IFU observations, and that the jet lobes corresponded spatially to kinematic components indicative of outflows or gas with increased turbulence. 

Frustrated radio jets or radiative AGN outflows are therefore likely to be important forces in galaxy evolution, especially since they appear to be commonplace in typical low luminosity Seyferts \citep{Gallimore2006}, which outnumber powerful radio galaxies at nearly ten to one. We note that the \citet{Gallimore2006} sample includes many objects with kiloparsec-scale radio structures that are likely related to star formation, so their occurrence rate of 44\% is understandably higher than our rate of 11\%. Our rate would be higher if we did not impose the criterion that the radio luminosity exceeds the far-IR radio correlation by a factor of 10 to be considered a jet. If we instead based our classification only upon jet-like one-sided or linear morphology, our occurrence rate of jet-like outflows would be $\sim22$\%. Including all of our kiloparsec-scale structures regardless of morphology, 41\% showed resolved radio emission. We further note that the \citet{Gallimore2006} parent AGN sample was selected in the optical and infrared, while ours was selected in the ultra-hard X-ray. 

The case of 2MASX~J0423+0408 is perhaps surprising. It has by far the largest jets in the sample, with lobes extending far outside the host galaxy, but resides squarely within the main sequence. One might expect that the star formation rate would be highly suppressed in such a target. This galaxy has been recently studied in great multi-wavelength detail by \citet{Fischer2019}, where we found that high gas velocity kinematics at large radii are likely due to mechanical driving from AGN winds impacting the ISM. The interaction shocks the ISM, producing thermal X-ray emission and cosmic rays, which in turn promote the formation of the observed S-shaped radio structure. The result is a pseudo-jet morphology along the high density ISM lanes. The object is also undergoing a merger, which may be enhancing the star formation (and may also have triggered the outflow).

If indeed the SFRs in our sample are being disrupted by the destruction or removal of the molecular gas, one might expect that the measured star formation rate would remain high for a few tens of Myr after the molecular gas is disturbed, before the signatures of star formation die away. Such a mechanism has been proposed to explain why some AGN samples with ionized gas outflows do not show any reduction in SSFR \citep{Woo2017}. The star formation efficiency (SFE) measures the rate of star formation versus the availability of molecular gas, SFR/$M_{\mathrm{H}_2}$ \citep[see the review by][]{Omont2007}. Having this information immediately to hand, we calculate the efficiencies for our sample using the SFRs in Table~\ref{t:tab1} and the molecular gas masses in Table~\ref{t:energy}. The values of SFE in our sample range from $-9.43$~yr$^{-1} < $log(SFE)$< -8.25$~yr$^{-1}$, which is consistent with the median range of normal late-type galaxies from \citet{Boselli2002}. Therefore, our SFEs are not particularly high, which may argue against the interpretation of molecular gas disruption or removal as the feedback mechanism.

Finally, we note that a jet expanding into the ISM is also capable of \emph{enhancing} the host star formation by creating pressure-driven gravitational instabilities in giant molecular clouds it encounters along the way \citep[e.g.,][]{Rees1989,Bicknell2000,Klamer2004,Croft2006}. The situation is quite complex, however, as shown in simulations by \citet{Antonuccio-Delogu2008}: jets with powers $10^{41} - 10^{47}$ erg~s$^{-1}$ can occasionally increase the star formation rate in clouds that they encounter, but the jet also causes fragmentation and establishes a difficult cooling environment for the clouds, which causes an overall inhibition of star formation.

\section{Conclusions}
\label{sec:conclusions} 
Out of our sample of 100 radio-quiet ultra-hard X-ray selected AGN, we find kiloparsec-scale radio structures not associated with star formation in 11 objects, usually on small physical scales compared to the host size (Figure~\ref{fig:hosts}). These structures can be interpreted either as small-scale AGN jets or as signatures of the interaction of a radiative AGN outflow with the host ISM.

We find that for 10 of the 11 objects with KSRs, the host galaxy star formation rate is suppressed to at least 1$\sigma$~ below the star forming main sequence, and in most cases much more, as shown in Figure~\ref{fig:sfms}. 

We estimate the energy indicated by the radio structures under the jet and outflow interpretations, and use these energies to evaluate two criteria for feedback: the ability to remove the CO-derived molecular gas mass from the galaxy and the kinetic energy transfer to molecular clouds leading to $v_\mathrm{cloud} > \sigma_*$.  

We conclude that KSRs related to AGN outflows or jets occur in at least 10\% of X-ray selected, radio-quiet Seyferts and have a significant effect on the star formation in their host galaxies, even when the physical scale of the radio emission is very small compared to the host galaxy extent. We further conclude that the star formation suppression is most likely due to the influence of a radiative AGN outflow that causes a pseudo-jet radio structure through ISM interactions, and not due to a scaled-down version of the relativistic jets seen in radio-loud AGN. 

Because these radio structures are common and occur in many more galaxies than the powerful radio jets typically associated with feedback, they are likely to be an important component of AGN-driven galaxy evolution in the universe.

\acknowledgments

Support for KLS was provided by the National Aeronautics and
Space Administration through Einstein Postdoctoral Fellowship
Award Number PF7-180168, issued by the Chandra
X-ray Observatory Center, which is operated by the
Smithsonian Astrophysical Observatory for and on behalf
of the National Aeronautics Space Administration
under contract NAS8-03060.  MK acknowledges support from NASA through ADAP award NNH16CT03C. CR acknowledges Fondecyt Iniciacion grant 11190831.

\small
\bibliographystyle{aasjournal}
\bibliography{biblio_bassjet}

 \end{document}